\documentclass[prx,aps,amssymb,onecolumn,noshowpacs,hyperref]{revtex4-1}
\usepackage{color}
\usepackage{graphicx}
\usepackage{hyperref}
\usepackage{amsmath}
\usepackage{latexsym}
\usepackage{amssymb}
\usepackage{mathrsfs}
\usepackage{mathtools}
\usepackage{layout}
\usepackage{verbatim}
\usepackage{multirow}
\usepackage{bm}
\usepackage{amsfonts,epsfig}
\usepackage{textcomp}
\usepackage{diagbox}
\usepackage[dvipsnames]{xcolor}
\definecolor{myblue}{RGB}{0, 100, 200}
\definecolor{myblue2}{RGB}{0, 100, 100}

\usepackage{geometry}
\usepackage{booktabs}
\usepackage{array}

\begin{document}

\title{Ancilla-assisted multi-qubit quantum gates with polar molecules}
\author{Yan Lu}
\affiliation{Center for Theoretical Physics and School of Physics and Optoelectronic Engineering, Hainan University, Haikou 570228, China}
\author{Xiao-Feng Shi}
\affiliation{Center for Theoretical Physics and School of Physics and Optoelectronic Engineering, Hainan University, Haikou 570228, China}
\date{\today}
\begin{abstract}

Quantum error correction enables reliable quantum computation using imperfect physical gates, but its implementation often requires multi-qubit controlled-unitary gates $\text{C}_{\mathbb{N}}(U)$ acting on $\mathbb{N} \ge 3$ qubits. Their direct realization is challenging, while decomposing them into standard one-qubit and two-qubit gates can increase circuit depth and error rates. Here, we present an ancilla-assisted scheme to directly implement $\text{C}_{\mathbb{N}}(U)$ in optically trapped ultracold polar molecules. By appropriately selecting qubit states for the ancilla and data qubits, dark states can appear during two smooth $\pi$ pulses. This can generate a controlled-phase gate with the phase set by the relative phase between the pulses; notably, a single $2\pi$ pulse yields a $\text{C}_{\mathbb{N}}(Z)$ gate. Numerical analysis for a simple molecular spatial configuration demonstrates that the gate infidelity is dominated by qubit decoherence, yielding a $\text{C}_2(Z)$ fidelity of $0.996$ estimated with $T_2^\ast=1.5$~s and a gate duration of $T \approx 4\text{~ms}$. Furthermore, diabatic errors remain low, on the order of $10^{-4}$ for $\mathbb{N}=2,3$ and $10^{-3}$ for $\mathbb{N}=4$, even with dipole-dipole interaction strengths as low as $25\hbar/T$.

\end{abstract}
\maketitle

\section{Introduction}
A key requirement for building a circuit-based fault-tolerant quantum computer is the capability to perform both high-efficiency information processing and quantum error correction. To process information, a universal set of quantum gates consisting of canonical one and two-qubit quantum gates is required~\cite{Williams2011}. Typical physical platforms can achieve a high fidelity $\mathcal{F}$ for them, such as $\mathcal{F} > 0.9999$ in two-qubit gates with trapped ions~\cite{hughes_trapped-ion_2025}. However, to correct unavoidable errors during computation, multi-qubit controlled gates $\text{C}_{\mathbb{N}}(U)$ are often required, where $\mathbb{N} \ge 3$ and $U$ is a one-qubit unitary operating on a target qubit~\cite{Crow2016,Petrosyan_2024}. Directly realizing such $\mathbb{N}$-qubit gate with a high fidelity is difficult.

A second prerequisite for fault-tolerant quantum computing is scalability. Solid-state architectures such as superconducting circuits~\cite{Castelvecchi_2023} and semiconductor quantum dots~\cite{Donnelly_2026} offer one path toward scaling, while neutral atoms trapped in optical tweezers present a scalable and reconfigurable alternative~\cite{Bluvstein2022}. Qubit arrays with over 6,000 neutral atoms were demonstrated~\cite{6100atoms}, and advanced optical tweezer techniques promise arrays of sub-million sites~\cite{Holman_2026}. With neutral atoms, Rydberg-mediated dipole-dipole interactions (DDI) enable two-qubit gates within the computational state space with gate fidelities up to $\mathcal{F} > 0.9994$~\cite{evered_high-fidelity_2026}. Nevertheless, these Rydberg gates typically rely on short-lived low-$l$ Rydberg states whose radiative lifetimes place a stringent upper bound on achievable gate fidelities~\cite{norrell_entangling_2026}. While long-lived high-$l$ Rydberg states could circumvent this issue, rapidly and coherently exciting them to meet the speed requirements of fault-tolerant architectures remains difficult~\cite{Cohen_2021,M_haignerie_2025,Machu_2026}. Meanwhile, mid-circuit error correction protocols are studied to mitigate operation errors, with circuit depths rapidly shooting up when the error per gate increases.

An alternative tweezer-based quantum computing platform is polar molecules~\cite{Cornish_2024}. Polar molecules possess multiple low-lying stable rotational states within their electronic ground manifold, offering an individually controllable Hilbert space for encoding information~\cite{Hepworth_2025}. Moreover, DDI can occur directly between polar molecules prepared in superpositions of these ground-manifold states. This contrasts sharply with neutral atoms, where DDI can occur only when excited to short-lived Rydberg states~\cite{norrell_entangling_2026}. Recent experiments have demonstrated two-molecule Bell states~\cite{Bao_2023,Holland_2023,ruttley_long-lived_2025,picard_entanglement_2025,yu2026} with fidelities up to 0.976~\cite{ruttley_long-lived_2025,yu2026}, as well as two-qubit $\text{iSWAP}$ gates with a fidelity of 0.92~\cite{picard_entanglement_2025}. While several theoretical schemes by using molecular DDI have been proposed for Bell-state and two-qubit gate generation~\cite{DeMille_2002,Zhu_2013,Yelin_2006,Ni_2018,Hughes_2020,Tscherbul_2023,bergonzoni_iswap_2025,Muminov_2026,Lu2026gate,Lu2026gate2}, it is unclear how to efficiently implement $\text{C}_{\mathbb{N}}$ with $\mathbb{N} \ge 3$ in polar molecules.

In this work, we propose an ancilla-assisted adiabatic dark-state protocol that efficiently implements $\text{C}_{\mathbb{N}}$ gates in ultracold polar molecules using only a single $2\pi$ pulse or a sequence of two smooth microwave $\pi$ pulses. A primary challenge in the study of polar molecules is the in-trap quantized motion~(QM)~\cite{Bao_2023,Holland_2023,ruttley_long-lived_2025,picard_entanglement_2025,yu2026}. By treating the molecular motion quantum-mechanically~\cite{Lu2026gate,Lu2026gate2}, we show that our ancilla-assisted $\text{C}_2(Z)$, where $Z$ is the third Pauli matrix, can suppress motion-induced infidelity to the order of $10^{-4}$, reaching a regime where gate performance is limited mainly by the qubit coherence time. We further evaluate $\text{C}_3(Z)$ and $\text{C}_4(Z)$ fidelity for a representative configuration of molecules' positions, finding that the dominant residual error arises from the breakdown of adiabaticity. Importantly, the diabatic error stems from unwanted DDI between data molecules and, hence, is not intrinsic but configuration-dependent. Diabaticity could be further minimized by tailoring smooth analytical pulse shapes~\cite{bergonzoni_iswap_2025,Lu_2026} or optimizing spatial configurations. Here, however, our focus is to establish the feasibility and efficiency of realizing multi-qubit gates with polar molecules.

While dark-state mechanisms have been explored extensively in, e.g., Rydberg-mediated quantum gates~\cite{Shi2021qst}, those implementations typically require shelving at least one atom on a Rydberg state before launching the dark-state sequence, introducing unavoidable Rydberg-state decay and Doppler dephasing~\cite{Saffman2011,Shi2020prapplied,Shi_2025-polar}. As a result, adiabatic Rydberg gates, though conceptually interesting~\cite{Shi2021qst}, have proven difficult to execute with high fidelity experimentally~\cite{McDonnell2022}. In contrast, our molecular dark-state scheme offers two  strengths: (1) all operational states belong to long-lived rotational manifolds, and (2) microwave driving of rotational transitions incurs virtually negligible Doppler dephasing compared to optical excitation of Rydberg atoms.

The remainder of this article is as follows. In Sec.~\ref{sec02}, we present the details to realize a two-qubit ancilla-assisted dark-state gate. In Sec.~\ref{sec03}, we take a level configuration with $^{23}$Na$^{133}$Cs to show the feasibility to realize the dark-state condition with polar molecules. In Sec.~\ref{sec04}, we study the gate errors from decoherence, imperfect adiabaticity, and wrong state transitions in the microwave pulses. In Sec.~\ref{sec05}, we study the gate error from the quantized in-trap molecular motion. In Sec.~\ref{sec06}, we analyze three-qubit and four-qubit controlled gates based on a simple spatial configuration of molecules. We give discussions in Sec.~\ref{sec07} and summarize in Sec.~\ref{sec08}.

\begin{figure}
\includegraphics[width=6.0in]
{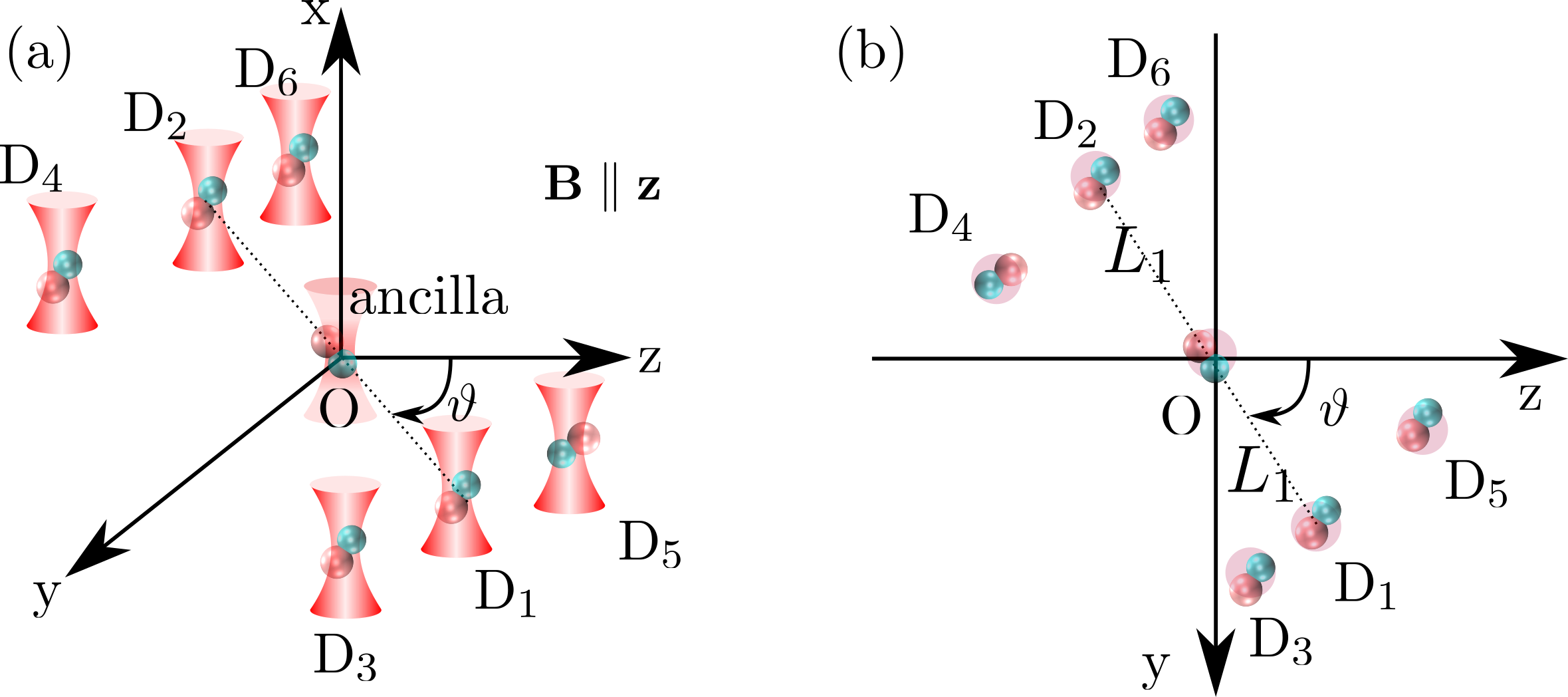}
\caption{Illustration of seven optically trapped polar molecules, where the one at $O(0,0,0)$ is the ancilla. The centers of the seven traps are in the y-z plane, with the tweezer light along $\bf{x}$. The angle between $\bf{z}$ and $\overrightarrow {D_{2}D_{1}}$ is equal to $\vartheta=$arccos$(1/\sqrt{3})$. The three sites $D_{j},D_{j+2},D_{j+4}$ are on one line for both $j=1$ and $j=2$, tilted from $\mathbf{z}$ by an angle $\vartheta$, too. For the two-qubit controlled-phase gate studied in Secs.~\ref{sec02},~\ref{sec03}, and~\ref{sec04}, the two data molecules are trapped at $D_1$ and $D_2$. For the three-qubit or four-qubit gates studied in Sec.~\ref{sec06}, the data molecules are trapped at $D_3,D_4,D_5$ and $D_6$. Here, a ``controlled-phase gate'' means a gate that leaves all the state components intact but renders a phase change $\varphi$ to the state component $\lvert \downarrow\downarrow\cdots\downarrow\rangle$, i.e., the state when each data qubit is initialised in $\downarrow$. 
\label{figure-1d} }
\end{figure}

\section{Ancilla-assisted dark-state gate in two molecules}\label{sec02}
We consider three polar molecules trapped at the three sites $O,~D_1$ and $D_2$ of Fig.~\ref{figure-1d}. The one at $O$ is an ancilla, labeled by ``a'', and the two at $D_{1,2}$ are data molecules, labeled c and t, where c and t are abbreviations of ``control'' and ``target''. For each data molecule, three states are involved, including two qubit states, $\lvert \uparrow\rangle$ and $\lvert\downarrow\rangle$, and an excited state $\lvert e\rangle$ that will mediate a resonant electric DDI in the absence of external electric field. To induce a quantum gate, we introduce an ancillary molecule initialised in the state $\lvert \downarrow_{\text{a}} \rangle$, where the three states $\uparrow_{\text{a}},\downarrow_{\text{a}}$ and $e_{\text{a}}$ differ from the three counterparts in the data qubit. A microwave field is sent to the ancilla which induces the Hamiltonian,
\begin{eqnarray}
\hat{H}_{\mu} = \hbar\frac{\Omega_{\mu}(t)}{2}\lvert \downarrow_{\text{a}}\rangle\langle e_{\text{a}}\rvert +\text{H.c.} ,\label{H_mu-1}
\end{eqnarray}
and only occurs for the ancilla; physical realization of this condition is shown later in Sec.~\ref{sec03}. We consider a condition that there will be a DDI in the state of the ancilla and one data molecule $\lvert e_{\text{a}},\uparrow\rangle$ or $\lvert \uparrow_{\text{a}},e \rangle$. This DDI can be written as $\hat{V} = \mathscr{J}  \lvert \uparrow_{\text{a}},e\rangle\langle e_{\text{a}},\uparrow\rvert+ \mathscr{J}^\ast   \lvert e_{\text{a}},\uparrow\rangle\langle \uparrow_{\text{a}},e\rvert$~\cite{picard_entanglement_2025}, where $\mathscr{J}$ is the DDI which can fluctuate. The state symbols in each ket for the ancilla are with subscript `a', while those for the data qubits are not with subscript for brevity.

 In the context of quantum logic gates, it is useful to study the gate map by examining the time dynamics for the eigenstates of the gate, which are $\lvert \downarrow,\downarrow\rangle, \lvert \downarrow,\uparrow\rangle,\lvert \uparrow,\downarrow\rangle,\lvert \uparrow,\uparrow\rangle$ here. In the three-molecule systems consisting of the two data molecules and the ancillary molecule, there are three different classes of input states:\newline (i) $ \lvert \downarrow_{\text{a}},\downarrow,\downarrow\rangle$;\newline (ii) $ \lvert \downarrow_{\text{a}},\downarrow,\uparrow\rangle$,$ \lvert \downarrow_{\text{a}},\uparrow,\downarrow\rangle$; \newline (iii) $ \lvert \downarrow_{\text{a}},\uparrow,\uparrow\rangle$.\newline Two smooth adiabatic pulses can implement a dark-state molecular gate. To show this, we consider a condition that the DDI between the ancilla and either data molecule is equal. For class (i) where no DDI will occur, the standard resonant Rabi oscillations will proceed. For class (ii), the state dynamics for $ \lvert \downarrow_{\text{a}},\uparrow,\downarrow\rangle$ and $ \lvert \downarrow_{\text{a}},\downarrow,\uparrow\rangle$ are similar, so that we can take, e.g., the former as an example, which evolves under the Hamiltonian
\begin{eqnarray}
\hat{H}_{\text{ii}} &=&\left(
\begin{array}{ccc}
0&\mathscr{J}& 0\\
 \mathscr{J}^\ast&0 & \hbar\frac{\Omega_\mu^\ast}{2} \\
 0 & \hbar\frac{\Omega_\mu}{2} & 0
\end{array}
\right)\label{adia-01}
\end{eqnarray}
in the basis of $ \{    \lvert \uparrow_{\text{a}},e,\downarrow\rangle      ,  \lvert e_{\text{a}},\uparrow,\downarrow\rangle     ,  \lvert \downarrow_{\text{a}},\uparrow,\downarrow\rangle  \} $. Equation~(\ref{adia-01}) can also be written as
\begin{eqnarray}
\hat{H}_{\text{ii}} &=& E\sum_{\beta=\pm}\beta  \lvert\beta\rangle\langle\beta\rvert,\label{adia-02}
\end{eqnarray}
where $E= \sqrt{|\mathscr{J}|^2+|\hbar\Omega_\mu|^2/4}$, and
\begin{eqnarray}
\lvert \pm\rangle &=& \frac{1}{ \sqrt{2}E }\left(\mathscr{J} \lvert \uparrow_{\text{a}},e,\downarrow\rangle      \pm E \lvert e_{\text{a}},\uparrow,\downarrow\rangle + \hbar\frac{\Omega_\mu}{2}  \lvert \downarrow_{\text{a}},\uparrow,\downarrow\rangle \right),
\label{adia-03}
\end{eqnarray}
where in principle Eq.~(\ref{adia-02}) should have three states, but during transferring from Eq.~(\ref{adia-01}) to Eq.~(\ref{adia-02}) there appears a zero-energy state, i.e., the dark state~\cite{Shi2021qst} which has an eigenenergy equal to zero, with an eigenstate given by
\begin{eqnarray}
\lvert 0\rangle &=& \frac{1}{E}\left(\hbar\frac{\Omega_\mu^\ast}{2} \lvert \uparrow_{\text{a}},e,\downarrow\rangle -\mathscr{J}^\ast  \lvert \downarrow_{\text{a}},\uparrow,\downarrow\rangle    \right).
\label{adia-04}
\end{eqnarray}

Consider a smooth pulse of duration $T$, where $\Omega_\mu$ goes to a maximal value $\Omega$, and then back to zero again. If the change of $\Omega_\mu$ is slow enough, then $\lvert \downarrow_{\text{a}},\uparrow,\downarrow\rangle$ evolves to $\lvert0\rangle$, and back to $\lvert \downarrow_{\text{a}},\uparrow,\downarrow \rangle$, without accumulating any phase because the eigenenergy for $\lvert 0\rangle$ is zero and no geometrical phase appears since the time evolution retraces. If we consider
\begin{eqnarray}
\pi&=& \int_0^T \Omega_\mu dt,\label{pi-pulse}
\end{eqnarray}
then the input state of class (i), $\lvert  \downarrow_{\text{a}},\downarrow,\downarrow\rangle$, evolves to $-i\lvert  e_{\text{a}},\downarrow,\downarrow\rangle$.

Soon after the above $\pi$ pulse, a second smooth pulse of equal pulse shape, but with a phase shift $\pi-\varphi$ is sent to the two molecules. According to the adiabatic picture shown above, the same dynamics according to Eqs.~(\ref{adia-01}),~(\ref{adia-02}),~(\ref{adia-03}), and~(\ref{adia-04}) will occur to the input state $\lvert \downarrow_{\text{a}},\uparrow,\downarrow\rangle$. After the second $\pi$ pulse, the states of class (ii) remain the same as the original ones. Because of the extra phase in the Rabi frequency during the second $\pi$ pulse, the input state $\lvert  \downarrow_{\text{a}},\downarrow,\downarrow\rangle$ for class (i), which is $-i\lvert  e_{\text{a}},\downarrow,\downarrow\rangle$ at the beginning of the second $\pi$ pulse, evolves to $e^{i\varphi}\lvert  \downarrow_{\text{a}},\downarrow,\downarrow\rangle$.

For the state in class (iii), the Hamiltonian is
\begin{eqnarray}
\hat{H}_{\text{iii}} &=&\left(
\begin{array}{cccc}
0& \mathscr{J}_{\text{ct}}& \mathscr{J}_{\text{ac}}&0\\
\mathscr{J}_{\text{ct}}^\ast&0& \mathscr{J}_{\text{at}}& 0\\
 \mathscr{J}_{\text{ac}}^\ast& \mathscr{J}_{\text{at}}^\ast &0& \hbar\frac{\Omega_\mu^\ast}{2} \\
 0 &0&  \hbar\frac{\Omega_\mu}{2} & 0
\end{array}
\right)\label{duu}
\end{eqnarray}
in the basis of $ \{   \lvert \uparrow_{\text{a}},e,\uparrow\rangle ,\lvert \uparrow_{\text{a}},\uparrow,e\rangle,  \lvert e_{\text{a}},\uparrow,\uparrow\rangle     ,  \lvert \downarrow_{\text{a}},\uparrow,\uparrow\rangle  \} $, where $\mathscr{J}_{\alpha\beta}$ is the DDI between the molecules labeled $\alpha,\beta\in\{$a,~c,~t$\}$. The reason that there can be a DDI for the states $\lvert \uparrow_{\text{a}},e,\uparrow\rangle$ and $\lvert \uparrow_{\text{a}},\uparrow,e\rangle$ is that if an electric DDI occurs for the states $\lvert e_{\text{a}},\uparrow\rangle$ or $\lvert \uparrow_{\text{a}},e \rangle$ between the ancilla and the data molecules, then the parities for the two states $e$ and $\uparrow$ of the data molecule should be opposite, so that there can be a resonant DDI that induces a state exchange $\lvert \uparrow_{\text{a}},e,\uparrow\rangle\leftrightarrow\lvert \uparrow_{\text{a}},\uparrow,e\rangle$.

To recover the dark-state condition for the state in class (iii), it is necessary to annul the DDI between the data molecules. This can be achieved by placing the molecules in a configuration so that the DDI between the two data molecules
\begin{eqnarray}
\mathscr{J}_{\text{ct}}&=&\mathbb{J}_{\text{ct}} \frac{1-3\cos^2\theta}{2} \frac{(2L_1)^3}{|\hat{\mathbf{r}}_{\text{c}}-\hat{\mathbf{r}}_{\text{t}}|^3}\label{J-fluctuation0-0}
\end{eqnarray}
vanishes when $1-3\cos^2\theta=0$, where $2L_1$ is the distance between the two data molecules at $D_1$ and $D_2$ as shown in Fig.~\ref{figure-1d}, $\mathbb{J}_{\text{ct}} =\frac{1 }{ 2\pi\epsilon_0(2L_1)^3}  \left(\frac{\text{\textdong}}{\sqrt{3}} \right)^2$ as shown in Appendix~\ref{app-ddi}, with \textdong~being the molecule-frame electric dipole moment, and $\theta$ is the angle between the quantization axis and the separation orientation of the control and target molecules. With the configuration of Fig.~\ref{figure-1d}, we have $\theta=$arccos$(1/\sqrt{3})$, so that $\mathscr{J}_{\text{ct}}$ becomes zero, and Eq.~(\ref{duu}) can be written as
\begin{eqnarray}
\hat{H}_{\text{iii}} &=&\left(
\begin{array}{ccc}
0&A& 0\\
 A&0 &\hbar \frac{\Omega_\mu^\ast}{2} \\
 0 &\hbar \frac{\Omega_\mu}{2} & 0
\end{array}
\right)\label{adia2-01}
\end{eqnarray}
in the basis of $ \{    \lvert \Lambda \rangle      ,  \lvert e_{\text{a}},\uparrow,\uparrow\rangle     ,  \lvert \downarrow_{\text{a}},\uparrow,\uparrow\rangle  \} $, where $A=\sqrt{|\mathscr{J}_{\text{ac}}|^2+|\mathscr{J}_{\text{at}}|^2 }$, and
\begin{eqnarray}
 \lvert \Lambda \rangle  &=& \frac{1}{A}\left( \mathscr{J}_{\text{ac}}\lvert \uparrow_{\text{a}},e,\uparrow\rangle +\mathscr{J}_{\text{at}}\lvert \uparrow_{\text{a}},\uparrow,e\rangle  \right).
\end{eqnarray}
Then, the dark-state picture for the input state $\lvert \downarrow_{\text{a}},\uparrow,\uparrow\rangle$ is recovered, and the two adiabatic pulses will not change any input state except of $\lvert  \downarrow_{\text{a}},\downarrow,\downarrow\rangle$ which picks up a phase factor $e^{i\varphi}$. Because the state of the ancilla is restored, the ancilla is again separated from the data qubits. Then, we have the state evolution for the qubit states,
\begin{eqnarray}
\lvert \uparrow,\uparrow\rangle&\mapsto& \lvert \uparrow,\uparrow\rangle,\nonumber\\
\lvert \uparrow,\downarrow\rangle&\mapsto& \lvert \uparrow,\downarrow\rangle,\nonumber\\
\lvert \downarrow, \uparrow\rangle&\mapsto& \lvert \downarrow, \uparrow\rangle,\nonumber\\
\lvert \downarrow,\downarrow\rangle&\mapsto& e^{i\varphi}\lvert \downarrow,\downarrow\rangle.\label{adia-5}
\end{eqnarray}
 For $\varphi=\pi$, one can see that a $2\pi$ microwave pulse of constant phase can realize the gate, which is the $\text{C}_2(Z)$ gate that can be transformed to a controlled-NOT via single-qubit rotations~\cite{Shi2021qst}.

\begin{figure}
\includegraphics[width=3.5in]
{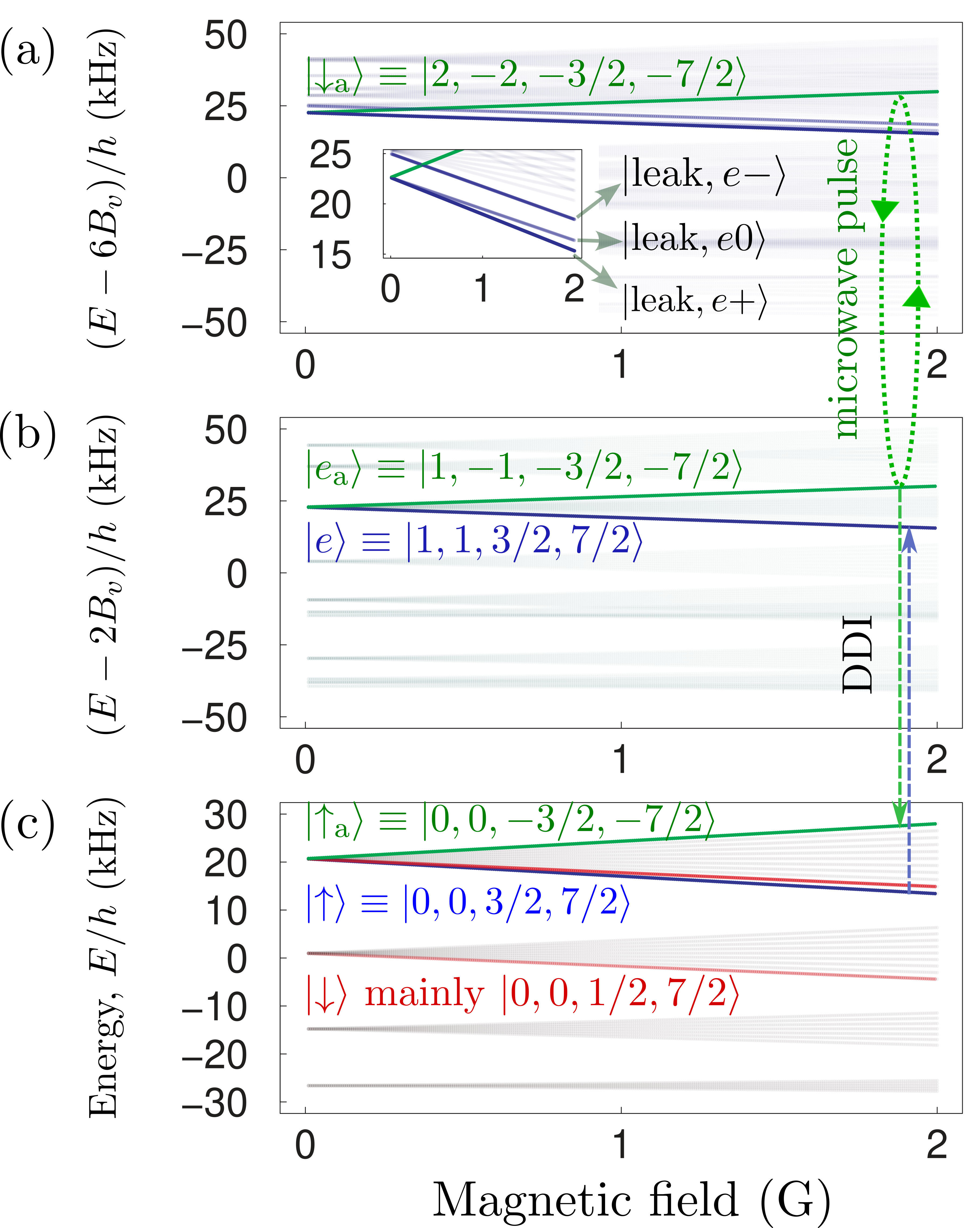}
\caption{(a,b,c) show the energy of the $N=2,~1$, and $N=0$ rovibrational ground manifold of $^{23}$Na$^{133}$Cs in zero electric field, where the energy in (a,~b) is shifted by $6B_v~(2B_v)$ with $B_v$ the rotational constant of the rovibrational ground manifold. The dark-state gate is realized by sending a $\sigma^-$-polarized microwave field resonant with the transition $\lvert\downarrow_{\text{a}}\rangle\leftrightarrow\lvert e_{\text{a}}\rangle$. The blue lines in (a) and its inset show $\lvert\text{leak},e0\rangle$ and $\lvert\text{leak},e\pm\rangle$, the leading leakage states that can be populated from $\lvert \text{e}\rangle$ by a $\pi$ and $\sigma^\pm$-polarized microwave field, respectively. As shown in Appendix~\ref{app-leakage}, $\lvert\text{leak},e+\rangle=\lvert2,2,3/2,7/2\rangle$, but both $\lvert\text{leak},e0\rangle$ and $\lvert\text{leak},e-\rangle$ have more than one state components due to hyperfine interaction. The transition induced by DDI denoted by the dashed arrows is resonant, i.e., the energy of the state $\lvert e_{\text{a}},\uparrow\rangle$ is equal to that of the state $\lvert \uparrow_{\text{a}}, e\rangle$.     \label{figure-hybrid} }
\end{figure}
\section{Physical realization}\label{sec03}
\subsection{Qubits with molecules of negligible rotational Zeeman effect}\label{sec03A}
The key to the dark-state gate of this article is a resonant DDI between the ancilla and each of the data molecules when the microwave field is resonant only with the ancilla. As illustrated in Fig.~\ref{figure-hybrid} calculated by Diatomic-py~\cite{Blackmore_2023}, with the $^1\Sigma^+$ ground manifold of $^{23}$Na$^{133}$Cs as an example, one can choose
\begin{eqnarray}
\lvert\uparrow_{\text{a}}\rangle &\equiv& \lvert0,0,-3/2,-7/2\rangle,~
\lvert e_{\text{a}}\rangle \equiv \lvert 1,-1,-3/2,-7/2\rangle,\nonumber\\
\lvert\uparrow\rangle &\equiv& \lvert0,0,3/2,7/2\rangle,~
\lvert e\rangle \equiv \lvert1,1,3/2,7/2\rangle.\label{phy-01}
\end{eqnarray}
Because the projections of the rotational angular momentum and the two nuclear spins onto the quantization axis are all maximal, there is no hyperfine-induced state mixing in the four states of Eq.~(\ref{phy-01}). So, the energy of each state in Eq.~(\ref{phy-01}) is given by $E=B_v N^2 -g_r\mu_{\text{\tiny{N}}} \mathbf{N}\cdot\mathbf{B}-\sum_{i\in\{\text{Na,~Cs} \}} g_i\mu_{\text{\tiny{N}}} \mathbf{I}_i\cdot\mathbf{B}(1-\sigma_i)  $~\cite{Aldegunde_2008,Aldegunde_2017}, where $B_v$ is the rotational constant, $\mathbf{N}$ is the rotational angular momentum, $g_r$ is the rotational g factor, $\{g_i,~\sigma_i\}$ are the nuclear spin g factor and shielding factor for nuclear spin $i\in~^{23}$Na,~$^{133}$Cs, where $I_{\text{\tiny{Na, Cs}}}$ are $3/2$ and $7/2$, respectively. One can see that if $g_r=0$, then the energy separation between $\lvert\uparrow_{\text{a}}\rangle $ and $\lvert e_{\text{a}}\rangle$ is equal to that between $\lvert\uparrow \rangle $ and $\lvert e\rangle$ because no state mixing occurs to them due to hyperfine interaction~\cite{PhysRevA.107.023102}. This is the case with $^{23}$Na$^{133}$Cs because until now, there is no report of a nonzero $g_r$ for $^{23}$Na$^{133}$Cs. 

Note that though $^{23}$Na$^{133}$Cs with $g_r=0$ is ideal, it is also possible to choose other molecules since the rotational Zeeman effect is orders of magnitude smaller compared to the nuclear spin Zeeman effect. Take $^{23}$Na$^{87}$Rb as an example, the value of $g_r$ is only $0.001(6)$, while $g_{\text{\tiny{Na(Rb)}}}(1-\sigma_{\text{\tiny{Na(Rb)}}} )$ is about 1.5~(1.8)~\cite{Guo_2018}. This means that if we choose similar states as Eq.~(\ref{phy-01}) when using $^{23}$Na$^{87}$Rb, the detuning for the DDI is of Hz scale with a Gauss-scale B-field. For DDI of several hundred Hz~\cite{picard_entanglement_2025}, a Hz-scale detuning is negligible, so that the dark-state condition remains.

Beside the four states of Eq.~(\ref{phy-01}), the other states that are involved in the gate can be $\lvert\downarrow_{\text{a}}\rangle = \lvert2,-2,-3/2,-7/2\rangle $ and $\lvert\downarrow\rangle $, where $\lvert\downarrow\rangle $ mainly contains $\lvert0,0,1/2,7/2\rangle$ but has multiple other components due to hyperfine interaction, with coefficients dependent on the applied magnetic field. If the theory in this work is used with molecules possessing a small $g_r$ the detuning in the DDI-coupled states can be neglected in a Gauss-scale magnetic field. Therefore, we take a B-field of 2 G as an example so as to be inclusive. At $B=$2~G, there is a state $\lvert\text{leak},e-\rangle$ from which the state $\lvert e\rangle$ can be populated, with a detuning $\Delta^{(\text{e-})}/h=-3.13$~kHz and a Rabi frequency $0.32\Omega_\mu$ as detailed in Appendix~\ref{app-leakage}, where $\Omega_\mu$ is the Rabi frequency of the transition $\lvert\downarrow_{\text{a}}\rangle\leftrightarrow\lvert e_{\text{a}}\rangle$. For a Rabi frequency around 1~kHz, this leakage can not be ignored, and a stronger B-field is needed to enlarge $\Delta^{(\text{e-})}$. For the example shown in Sec.~\ref{sec04}, this leakage error is orders of magnitude smaller than other errors, and, hence, can be ignored.

It was experimentally demonstrated that three states with maximally stretched spins in the $N=0,1,2$ rotational manifold can be simultaneously trapped by a magic-wavelength trap~\cite{Hepworth_2025}. Because states with positive or negative stretched spins experience equal light shifts~\cite{gregory2026}, the states $\lvert N,  N,3/2,7/2\rangle$ and $\lvert N, -N,-3/2,-7/2\rangle$ can be trapped together. So, by the same technique as in Ref.~\cite{Hepworth_2025}, it is possible to trap the states used in this article.

\subsection{DDI between molecules}\label{sec03B}
The DDI between two data molecules at $D_{i}$ and $D_j$ separated by a distance of $R_{ij}$ is the usual exchange interaction that swaps the rotational quantum states of the two molecules. As detailed in Appendix~\ref{app-ddi}, the interaction is given by
$V_{ij}= \frac{1}{4\pi\epsilon_0R_{ij}^3}\frac{\text{\textdong}^2}{6}(3\cos^2\theta-1)$, where \textdong~is the molecule-frame dipole moment, $\theta$ is the angle between $\mathbf{z}$ and the separation axis of the two molecules, and $\epsilon_0$ is the dielectric permittivity in free space. There is a factor $-1/2$ different from the usual DDI in the form of $\frac{1}{4\pi\epsilon_0R_{ij}^3}\frac{\text{\textdong}^2}{3}(1-3\cos^2\theta)$ due to the angular momentum coupling.

The DDI between the ancilla at $O$ and a data molecule at $D_j$ involves the total change of angular momentum $\pm2\hbar$, with $j=1,\cdots,6$, and Appendix~\ref{app-ddi} shows that the interaction is $V_{Oj}=\frac{\text{\textdong}^2}{4\pi\epsilon_0R^3}  \frac{1}{2}\sin^2\theta e^{-2i\phi}$, where $\phi$ is the azimuthal angle of the two-molecule separation axis and $R$ is the distance between the molecules.

\begin{table}[ht]
  \centering
\caption{Error budget of the ancilla-assisted $\text{C}_2(Z)$ gate with $|\mathscr{J}|/h=1$~kHz and $T\approx4.0$~ms. The leakage transition with pure polarization is too small, as shown in Sec.~\ref{sec04B0}, and, hence, not listed here. The infidelity from the polarization error is estimated by assuming an experimentally feasible $\varsigma_0\sim0.001$, and further by $\varsigma=100$ because the gate error is larger with a larger $\varsigma$ for each $\varsigma_0$. The error from the molecular motion is estimated by assuming $\ell_x/L_1=0.078$ and initial motional state $(\hat{a}_{x1}^\dag \hat{a}_{x2}^\dag )^2 \lvert \text{vac}\rangle/2$ as detailed in Sec.~\ref{sec0qmo}, where $\hat{a}_{x1}^\dag$ and $ \hat{a}_{x2}^\dag$ are the creation operators of two hybrid motional modes along the direction of the tweezer light. The decoherence error contributes $77\%$ to the total error here, due to the gate time is long compared to the assumed coherence time $T_2^\ast=1.5~$s here.    \label{table-fid}  }
  \begin{tabular}{cccc|c}
    \hline  \hline
 Decoherence  & Diabatic error  & Polarization impurity  &  Molecular motion & \multirow{2}{*}{ Total error} \\

 (Sec.~\ref{sec03C}) &(Sec.~\ref{sec04A}) & ( Sec.~\ref{sec04B}) &  (Sec.~\ref{sec0qmo}) & \\
     \hline
$2.7\times10^{-3}$ &   $1.96\times10^{-4}$     &  $1.91\times10^{-4}$    &  $4.13\times10^{-4}$ & $3.5\times10^{-3}$  \\
\hline
   \hline
  \end{tabular}

  \end{table}

\section{Infidelity of two-qubit gates}\label{sec04}
The realization of the dark-state mechanism relies on the adiabatic condition, and the state choice of Sec.~\ref{sec03} relies on circularly polarized microwave field to selectively excite the ancilla state transition, while impure polarization can incur leakage. Besides, there can be decoherence in the states, which can also lead to error. We analyze these with the example of a $\text{C}_2(Z)$ gate.

\subsection{Decoherence of the qubits}\label{sec03C}
Though the states involved in the gate can be trapped by the same trap with vanishing differential light shift in theory~\cite{gregory2026,Hepworth_2025}, there are still noise that hampers the coherence. When the maximally stretched molecules with rotational quantum number $N=0,1,2$ are trapped in the magic-wavelength tweezers as demonstrated in Ref.~\cite{Hepworth_2025}, the noise in the applied magnetic field can limit the coherence of the states in a much more dramatic way compared to other noise like the residual electric field. At the field of 181.7~G and for the transition between the $N=2$ and $N=0$ manifolds, the noise about 10~mG in the B-field limits the coherence to $T_2^\ast\approx2$~s in the specific apparatus of Ref.~\cite{Hepworth_2025}. This 2~s coherence time is the shortest among the transitions between the three stretched states with $N=0,1$, and $2$ of Ref.~\cite{Hepworth_2025}. For the transition in this article between the $N=1$ and $N=2$ states, Table~2 of Ref.~\cite{Hepworth_2025} indicates that the sensitivity coefficients would be about two times smaller with the magnetic noise, or five times smaller with the electric noise compared to the transition between $N=2$ and $N=0$. This means that if a strong magnetic field as that of Ref.~\cite{Hepworth_2025} is used, the coherence time here can be two times longer here. On the other hand, because the fluctuation of B-field scales with the base value of the B-field, when a Gauss-scale magnetic field is used, the magnetic noise can be much suppressed, and the coherence time can be greatly increased. This means that in principle, the coherence time of this work can be much longer.

When the three manifolds $N=0,1,2$ are involved, Ref.~\cite{Hepworth_2025} showed that the coherence time is larger than $1.5~$s for all the transitions among them. In this article, when the dark-state is involved, the transition from $N=2$ to $N=0$ state is at work, which is true for states in classes (ii) and (iii); but no dark state appears as for the state in class (i), so the transition is between the $N=2$ and $N=1$ manifolds. To be conservative, we estimate that the gate error from the decoherence to be $T/T_2^\ast$ with $T^\ast=1.5$~s, where $T$ is the gate duration.


The gate duration $T$ depends on the strength of DDI. To estimate DDI, we consider the three molecules trapped at the three sites $D_{1,2}$ and $O$ of Fig.~\ref{figure-1d}, where the ancilla is at $O$, and the control and target molecules are at $D_{1,2}$. The three molecules are on the $yz$ plane, and at angle $\vartheta=$arccos$(1/\sqrt{3})$ with respect to the z-axis. The interaction between the ancilla and the control or target involves change of total angular momentum of $\pm2$, and can be calculated to be $\frac{\text{\textdong}^2}{4\pi\epsilon_0L_1^3}  \frac{1}{2}\sin^2\vartheta e^{-2i\phi}$ as detailed around Eq.~(\ref{ancillaDDI}), where $L_1$ is the distance between the ancilla and each of the two data molecules. With $\phi=0$, we have $-\mathscr{J}=\mathbb{J}\equiv\frac{1}{3}\frac{\text{\textdong}^2}{4\pi\epsilon_0L_1^3}$, which is used in Eq.~(\ref{adia-01}). In Eq.~(\ref{adia2-01}), both $\mathscr{J}_{\text{ac}}$ and $\mathscr{J}_{\text{at}}$ are equal to $\mathscr{J}$ because the two sites $D_{1,2}$ are separated from $O$ by the same distance. Then, with $|OD_1|\approx1.29~\mu$m, one can have $\mathbb{J}/h=1$~kHz and $T\approx4$~ms as shown later in Eq.~(\ref{omega-time}). This leads to a decoherence error $2.7\times10^{-3}$ for the gate.
\begin{figure}
\includegraphics[width=5.5in]
{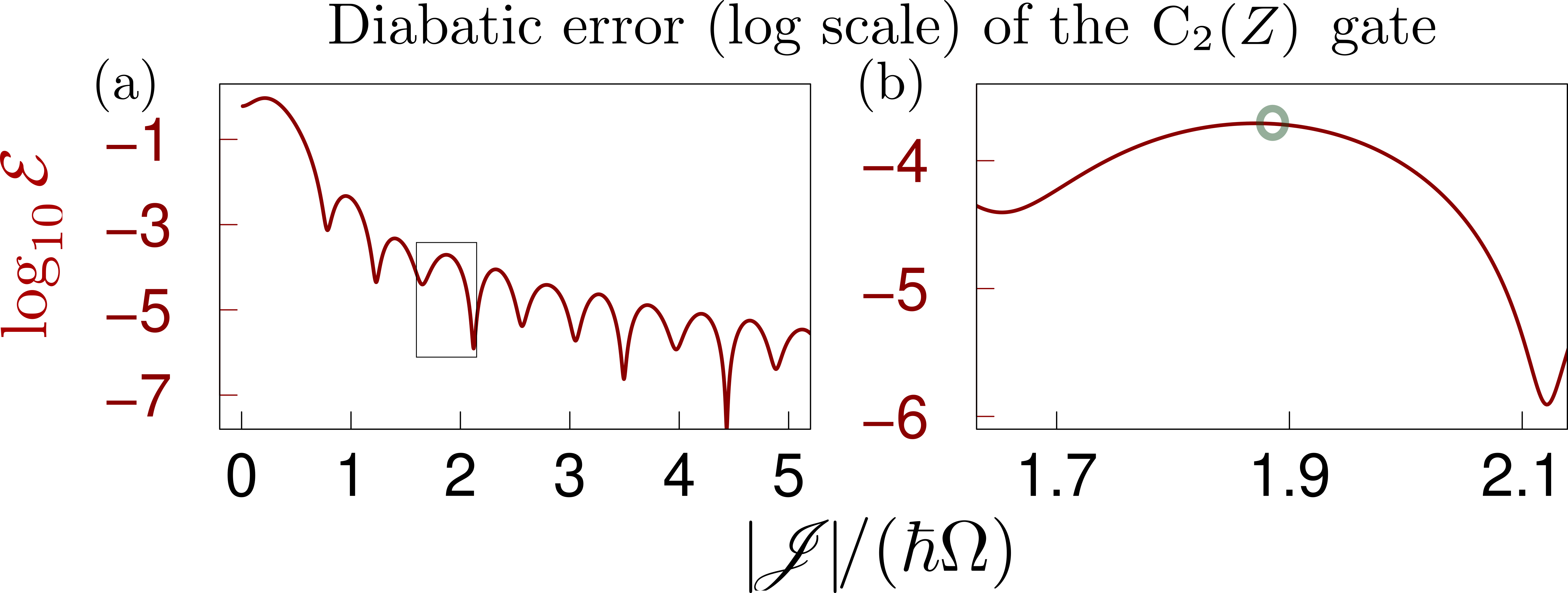}
\caption{The curve shows log$_{10}(1-\mathcal{F})$ of the $\text{C}_2(Z)$ gate simulated with the pulse in Eq.~(\ref{omega-time}), with $|\mathscr{J}|/(\hbar\Omega)\in(0,5)$ in (a) and $\in(1.7,2.1)$ in (b). In order to reveal the diabatic effect, the simulation assumes no decoherence of the two qubit states nor technique noise, but only accounts for the intrinsic rotation errors as compared to the desired dark-state mechanism detailed in Sec.~\ref{sec02}. There is some oscillation of the infidelity when varying $\mathscr{J}$ which is due to the detuned nature of the Rabi cycles in the presence of DDI~\cite{Shi2018prapp2,Lu2026gate}. In (a), one can see that with DDI comparable to the maximal of the Rabi frequency, the diabatic leakage can be small. The part enclosed by the rectangle in (a) is enlarged in (b), where the circle indicates $|\mathscr{J}|/(\hbar\Omega)\approx1.87$, at which $\log_{10}(1-\mathcal{F})$ has a local maximal, equal to $-3.71$. (b) shows that with a small deviation of $|\mathscr{J}|$ around $|\mathscr{J}|/(\hbar\Omega)\approx1.87$, the fidelity increases.
\label{figure-uni-error} }
\end{figure}
\subsection{Diabatic error}\label{sec04A}
Any gate based on adiabatic pulses is subject to diabatic leakage because it is impossible to realize the pulse infinitely slowly. The error due to the diabatic effect can be examined by varying the pulse duration. As shown above, if we want to realize a $\text{C}_2(Z)$ gate, i.e., $\varphi=\pi$ in Eq.~(\ref{adia-5}), then the two $\pi$ pulses can have equal phase as shown between Eq.~(\ref{pi-pulse}) and Eq.~(\ref{duu}). In this case, one $2\pi$ pulse can realize the $\text{C}_2(Z)$ gate, and we can consider a Gaussian-like pulse
\begin{eqnarray}
  \Omega_\mu(t) &=&\Omega \left(e^{-\frac{(t-T/2)^2}{2t_{\text{w}}^2}}-e^{-\frac{T^2}{8t_{\text{w}}^2}}\right),
  \label{omega-time}
\end{eqnarray}
where we set $t_{\text{w}}=0.234T$ at which the peak value of the Rabi frequency is smallest, equal to about $\Omega=13.5/T$, but meanwhile the pulse area is still $2\pi$~\cite{Lu2026gate}. The pulse in Eq.~(\ref{omega-time}) is not Gaussian because it starts and ends at zero, but has a finite duration $T$. In principle, pulse shaping~\cite{bergonzoni_iswap_2025,Lu_2026} can be applied to optimize the adiabatic effect with a finite pulse.

In Fig.~\ref{figure-uni-error}(a) and (b), we show the infidelity $1-\mathcal{F}$ of the $\text{C}_2(Z)$ gate in the log scale as a function of DDI in units of $\hbar\Omega$, where the fidelity $\mathcal{F}$ is defined according to Ref.~\cite{Pedersen2007} as detailed in Appendix~\ref{app-fidelity}. Remarkably, even with $|\mathscr{J}|$ comparable to $\hbar\Omega$, the infidelity from the diabatic effect can be quite small. For example, at $|\mathscr{J}|/(\hbar\Omega)=1.23$, $1-\mathcal{F}$ is only $4.4\times10^{-5}$. However, a small variation of DDI will push the error to the order of $10^{-3}$ at this point. To be robust against QM, we choose the circled point of Fig.~\ref{figure-uni-error}(b) with $|\mathscr{J}|/(\hbar\Omega)=1.87$ for the analysis later on because such a point has a locally minimal fidelity. The data for plotting Fig.~\ref{figure-uni-error}(a) shows that when $|\mathscr{J}|/(\hbar\Omega)>1.993$, all the fidelities are larger than $0.9999$, revealing that the ancilla-assisted dark-state evolution is quite robust even with a small $|\mathscr{J}|/(\hbar\Omega)$.

At the circled point of Fig.~\ref{figure-uni-error}(b), the average error for the ancilla state $\lvert\uparrow_{\text{a}}\rangle$ to be restored is $p_{\text{a}}=2.0\times10^{-4}$. This is mainly because in the case of the input state in class (ii), there is a population $3.7\times10^{-4}$ left in $\lvert e_{\text{a}}\rangle$ for each state in class (ii); for the state in class (iii), the population loss is drastically different, in both $\lvert e_{\text{a}}\rangle$ and $\lvert\downarrow_{\text{a}}\rangle$, with a total of $1.95\times10^{-5}$. Therefore the error $p_{\text{a}}$ can accumulate along the circuit unless re-initialization is used for it. However, note that if stronger DDI is used, this error can be suppressed.

\subsection{Leakage transitions}\label{sec04B0}
A potential error in the gate is leakage transitions~\cite{Ni_2018}. For the input states in class (ii) and (iii), the state $\lvert e\rangle$ in the data molecules can be transiently populated due to the DDI. When the microwave field is ideally $\sigma^-$ polarized, there is one leakage transition, $\lvert e\rangle\leftrightarrow\lvert \text{leak},e-\rangle$, which is detuned by about -3.13~kHZ at $B=2$~G. Due to hyperfine interaction, $\lvert \text{leak},e-\rangle$ has several components unless very strong B-field is applied. At $B=2$~G, the two major components are $\lvert2,0,3/2,7/2\rangle$ and $\lvert2,1,3/2,5/2\rangle$, with populations 0.597 and 0.245, respectively. More details on the analyses of the leakage transitions are grouped in Appendix~\ref{app-leakage}. With the detailed Hamiltonian including the leakage, we find that the extra error from this leakage transition is about $10^{-7}$. It is orders smaller than other error, and is not listed in Table~\ref{table-fid}.

\begin{figure}
\includegraphics[width=5.5 in]
{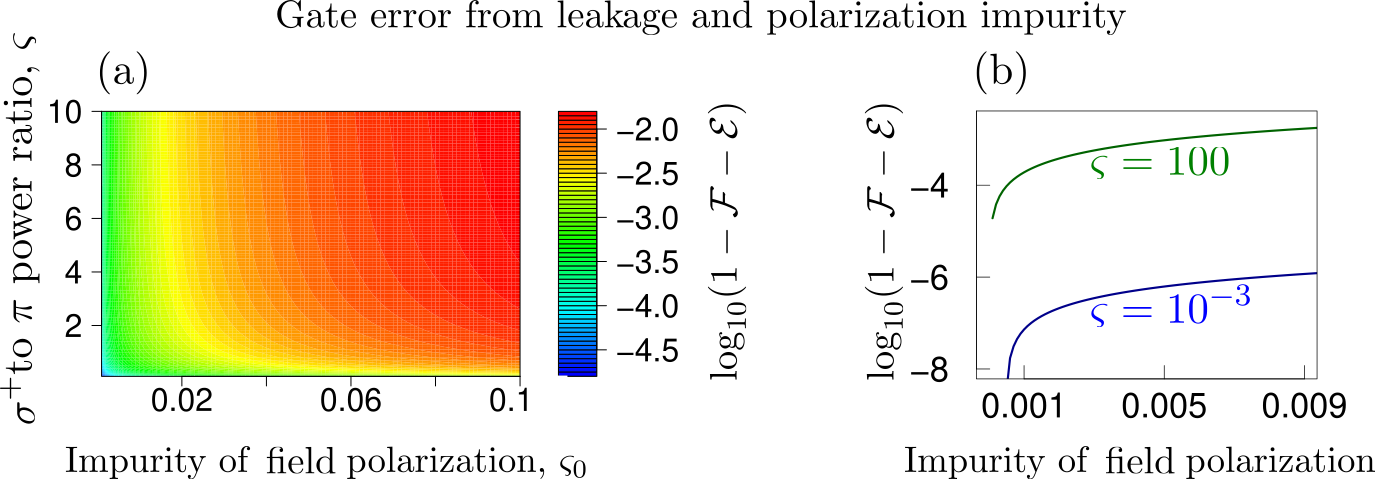}
\caption{Logarithm of the infidelity of the ancilla-assisted $\text{C}_2(Z)$ gate caused by leakage and impure polarization in the microwave field with the choice of states in Fig.~\ref{figure-hybrid} at $B=2$~G and $|\mathscr{J}|/h=1$~kHz, which corresponds to $|OD_1|\approx1.29~\mu$m. $\mathcal{E}$ is the infidelity calculated without any impure polarization, i.e., is given in Fig.~\ref{figure-uni-error}. Here $\varsigma_0\in(10^{-3},0.1)$ is the intensity ratio of the wrong field to the desired field. (a) shows results with $\varsigma\in(0.1,10)$, where $\varsigma$ is the intensity ratio of the $\sigma^+$ to the $\pi$ polarized fields, shown in Eq.~(\ref{omega-impure}), and (b) shows results with $\varsigma=0.001$ and $100$. The peak value of $\Omega_\mu$ in Eq.~(\ref{omega-time}) is $|\mathscr{J}|/(1.87\hbar)$ with $T\approx4.0$~ms, which corresponds to the hollow circle of Fig.~\ref{figure-uni-error}(b) where the diabatic error reaches a local maximal $\mathcal{E}\approx1.96\times10^{-4}$. This figure shows that larger $\varsigma_0$ leads to larger gate error. With a given $\varsigma_0$, a larger $\varsigma$ yields a larger error, shown in (b). 
\label{figure-impure} }
\end{figure}
\subsection{Polarization impurity in the field}\label{sec04B}
When there is polarization impurity in the microwave field, more leakage can occur. As relevant to the input states in class (ii) and (iii), the state $\lvert e\rangle$ in the data molecules can be populated. Then, it can be excited to $\lvert \text{leak},e0\rangle$ with a $\pi$ polarized field, and to $\lvert \text{leak},e+\rangle$ with a $\sigma^+$ polarized field, where the former is detuned by -1.06~kHz and the latter is resonant. $\lvert \text{leak},e+\rangle$ is resonant because it is $\lvert2,2,3/2,7/2\rangle$, which is separated from $\lvert e\rangle$ by exactly the same energy as between $\lvert e_{\text{a}}\rangle$ and $\lvert\uparrow_{\text{a}}\rangle$. This resonant leakage occurs only when there is $\sigma^+$ component in the field. These three leaking states are denoted in the inset of Fig.~\ref{figure-hybrid}(a)

Impure polarization can also influence the input state of class (i), $\lvert\downarrow_{\text{a}}, \downarrow,\downarrow\rangle$. Note that for this input state, no DDI occurs, but when the microwave fields excites it to $\lvert e_{\text{a}}, \downarrow,\downarrow\rangle$, the microwave field can excite $\lvert e_{\text{a}} \rangle$ of the ancilla to states other than the state $\lvert\downarrow_{\text{a}} \rangle$. As detailed in Appendix~\ref{app-ea-leak}, there are two states in the $N=2$ manifold that can open the leakage channel for $\lvert e_{\text{a}} \rangle$ of the ancilla, one detuned by $ 1.04$~kHz, and the other by $- 1.54~\text{kHz}$. The Hamiltonian for the major leakage transitions are detailed in Appendix~\ref{app-leak-H}.

There are some states that seem to support leakage but can be ignored. As shown in Fig.~\ref{figure-hybrid}, the microwave field is for a transition between $N=1$ and $N=2$, so that the states $\lvert\downarrow\rangle$, $\lvert\uparrow_{\text{a}}\rangle$, and $\lvert\uparrow\rangle$ with $N=0$ are far detuned, with a detuning of order of $2B_v$, and, hence, do not cause leakage. For the input states in class (ii) and (iii), the state $\lvert e_{\text{a}}\rangle$ is barely populated because according to the dark-state picture outlined around Eq.~(\ref{adia-04}), the state $\lvert e_{\text{a}}\rangle$ is in the bright states $\lvert\pm\rangle$ defined in Eq.~(\ref{adia-03}). Due to the large energy gap between the dark state and the two bright states, the molecular state follows after the dark state. Therefore any leakage from $\lvert e_{\text{a}}\rangle$ is a higher-order process compared to the diabatic leakage. Since the diabatic leakage is studied in Sec.~\ref{sec04A}, we can ignore the higher-order leakage from $\lvert e_{\text{a}}\rangle$ for the states in classes (ii) and (iii).

With the three-molecule layout as employed in Sec.~\ref{sec04A}, we numerically study the infidelity from the field impurity, with the results shown in Fig.~\ref{figure-impure} as a function of the total power ratio $\varsigma_0$ of the wrong polarization, and the power ratio $\varsigma$ between the $\sigma^+$ and $\pi$ polarizations. Here, the ratio between the powers of the three polarizations is
\begin{eqnarray}
P(\sigma^-):P(\mathbf{z}):P(\sigma^+) &=&1:\frac{\varsigma_0}{1+\varsigma  }: \frac{\varsigma_0\varsigma}{1+\varsigma  }.
  \label{omega-impure}
\end{eqnarray}
We have used $|\mathscr{J}|/h=1$~kHz in Fig.~\ref{figure-impure} that can be realized by setting the ancilla-data molecule distance to be $|OD_1|\approx1.29~\mu$m. To solely show the effect of the leakage, we use a weak magnetic field $B=2$~G so that the detunings of the leakage transitions are small. The infidelity of Fig.~\ref{figure-impure} has deducted the error from the diabatic effect. One can see that the smallest fidelity is with the largest $\varsigma_0$ and $\varsigma$. Interestingly, we find that with a large $\varsigma_0=10\%$ but if $\varsigma=0$, there is nearly no extra infidelity beside the error from the diabatic leakage. But if $\varsigma_0=10\%$ and $\varsigma=10$, the gate fidelity is only $98.0\%$. This means that the major error when leakage occurs is from the transition $\lvert e\rangle\leftrightarrow\lvert \text{leak},e+\rangle$. This resonant leakage transition occurs once there is undesired $\sigma^+$ polarized field, leading to a larger error compared to other leakage transitions. Experimentally, it is possible to realize $\varsigma_0\sim0.001$, at which we find that the extra error beside the diabatic error is $1.91\times10^{-4}$ even with $\varsigma=100$. Thus, the dark-state gate with the level example of Fig.~\ref{figure-hybrid} can attain a high accuracy.

\section{DDI-QM coupling in the $\text{C}_2(Z)$ gate}\label{sec05}
The dark-state molecular gate in Eq.~(\ref{adia-5}) can attain a high fidelity in the presence of molecular motion. To study this, in principle we can preserve the motional states along the three directions. Then, each molecule has three vibrational modes. Together, this leads to nine modes for the three molecules at $O$, $D_1$ and $D_2$ in Fig.~\ref{figure-1d}. In Refs.~\cite{Lu2026gate,Lu2026gate2}, a mode separation technique was used, revealing that half of the modes are involved in a two-molecule system. Following Refs.~\cite{Lu2026gate,Lu2026gate2}, we first expand the DDI with small position fluctuation of molecules so as to reach a compact form of the DDI, as detailed in Appendix~\ref{app-expansion}. Then, Appendix~\ref{sec-mode-separate} shows that for the three-molecule case, there are nine modes, but only six are involved in the gate, with the remaining three decoupled.

\subsection{DDI in terms of bosonic operators }
The DDI-QM coupling among the three molecules is detailed in Appendix~\ref{sec-mode-separate}. As shown in Eq.~(\ref{H-trap01}), the Hamiltonian of the traps for molecule a, c, and t at $O$ and $D_{1,2}$ can be written as
\begin{eqnarray}
 \hat{H}_{\text{trap}} &=&  \hbar \sum_{\alpha=0}^2\sum_{\xi\in\{x,y,z\}}\omega_\xi \left(\hat{a}_{\xi\alpha }^\dag \hat{a}_{\xi\alpha}   +\frac{1}{2}\right),
\end{eqnarray}
where
\begin{eqnarray}
\hat{a}_{\xi\alpha} &=&  \frac{1}{\sqrt{2 } \ell_\xi}\left(\xi_{\alpha}+ \ell_\xi^2 \frac{\partial}{\partial \xi_{\alpha}}\right),
\hat{a}_{\xi\alpha}^\dag=  \frac{1}{\sqrt{2 } \ell_\xi} \left(\xi_{\alpha}- \ell_\xi^2 \frac{\partial}{\partial \xi_{\alpha}}\right). \label{a-b-mode-main}
\end{eqnarray}
Here, $\xi_\alpha$ are linear, orthogonal superposition of small deviation of the molecular coordinate away from the center of the trap for $\xi\in\{x,y,z\}$ as detailed in Eq.~(\ref{transform}), $\omega_{\xi}$ is the angular frequency of the trap frequency, and $\ell_\xi =  \sqrt{ \hbar/(m\omega_\xi)}$. For brevity, we assume that the three traps have the same trap frequency along each direction. For the configuration in Fig.~\ref{figure-1d}, the three molecules at $O$ and $D_{1,2}$ are along a line tilted away from $\mathbf{z}$ by $\vartheta=$arccos$(1/\sqrt{3})$. Then, the DDI between two molecules in each pair (a and c), (a and t), and (c and t) are
\begin{eqnarray}
\mathscr{J}_{\text{ac}} &\approx &  \frac{C_{\text{ac}}e^{-2i\phi}}{L_1^3}  \left[ \sin^2\vartheta+ \frac{   A_{y,1} y_{\text{ac}}  + A_{z,1} z_{\text{ac}}   } {L_1}  + \frac{A_{xx,2} x_{\text{ac}}^2  + A_{yy,2}y_{\text{ac}}^2+ A_{zz,2} z_{\text{ac}}^2 + A_{yz,2} y_{\text{ac}}z_{\text{ac}}}{L_1^2} \right] ,\nonumber\\
\mathscr{J}_{\text{at}} &\approx &  \frac{C_{\text{at}}e^{-2i\phi}}{L_1^3}  \left[ \sin^2\vartheta+ \frac{   A_{y,1} y_{\text{at}}  + A_{z,1} z_{\text{at}}   } {L_1}  + \frac{A_{xx,2} x_{\text{at}}^2  + A_{yy,2}y_{\text{at}}^2+ A_{zz,2} z_{\text{at}}^2 + A_{yz,2} y_{\text{at}}z_{\text{at}}}{L_1^2} \right] ,\nonumber\\
\mathscr{J}_{\text{ct}} &\approx & \frac{C_{\text{ct}}}{(2L_1)^3}  \left[  \frac{   A_{y,1}^\prime y_{\text{ct}}  + A_{z,1} ^\prime z_{\text{ct}}   } {2L_1}  + \frac{A_{xx,2}^\prime  x_{\text{ct}}^2  + A_{yy,2}^\prime y_{\text{ct}}^2+ A_{zz,2}^\prime  z_{\text{ct}}^2 + A_{yz,2}^\prime  y_{\text{ct}}z_{\text{ct}}}{(2L_1)^2} \right] ,\label{Jact-main}
\end{eqnarray}
where $C_{\text{ac}} =C_{\text{at}} =\frac{1}{4\pi\epsilon_0}\frac{\text{\textdong}^2}{2}$ and $C_{\text{ct}}= -\frac{1}{4\pi\epsilon_0}\frac{\text{\textdong}^2}{6}$ as shown in Appendix~\ref{app-ddi}, the coefficients $A_{\alpha,1},~A_{\alpha\beta,2}$ with $\alpha\in\{x,y,z\}$ are derived in Appendix~\ref{app-expansion}, and are given in Table~\ref{sensi-table}, $A_{\alpha,1}^\prime,~A_{\alpha\beta,2}^\prime$ are given in Table~\ref{sensi-table2}~(with $^\prime$ omitted), and
\begin{eqnarray}
x_{\text{ct}}&=&\ell_x (\hat{a}_{x2}^\dag + \hat{a}_{x2}),~ y_{\text{ct}} =   \ell_y (\hat{a}_{y2}^\dag + \hat{a}_{y2}), ~z_{\text{ct}}  =   \ell_z (\hat{a}_{z2}^\dag + \hat{a}_{z2}),\nonumber\\
 x_{\zeta}&=&  \frac{\ell_x}{2}\left[ \sqrt{3}  (\hat{a}_{x1}^\dag + \hat{a}_{x1}) +   \Upsilon_\zeta (\hat{a}_{x2}^\dag + \hat{a}_{x2}) \right], ~
y_{\zeta}=   \frac{\ell_y}{2}\left[ \sqrt{3}  (\hat{a}_{y1}^\dag + \hat{a}_{y1}) +  \Upsilon_\zeta  (\hat{a}_{y2}^\dag + \hat{a}_{y2}) \right],\nonumber\\
z_{\zeta}&=&   \frac{\ell_z}{2}\left[ \sqrt{3}  (\hat{a}_{z1}^\dag + \hat{a}_{z1}) +  \Upsilon_\zeta  (\hat{a}_{z2}^\dag + \hat{a}_{z2}) \right],\label{main-ac-03}
\end{eqnarray}
where $\Upsilon_\zeta=1(-1)$ if $\zeta=$at~(ac). In Eq.~(\ref{Jact-main}), the phase term in $\mathscr{J}_{\text{ac}}$ and $\mathscr{J}_{\text{at}}$ are given by
\begin{align}
e^{-2i\phi} \approx  -1 - \frac{2i}{L_1\sin\vartheta} x_\zeta + \frac{2}{(L_1\sin\vartheta)^2}x_{\zeta}^2 + \frac{2i}{(L_1\sin\vartheta)^2} x_{\zeta} y_{\zeta},\label{main-phi-expansion2}
\end{align}
with $\zeta$ = ac or at.

\begin{figure}
\includegraphics[width=6.0in]
{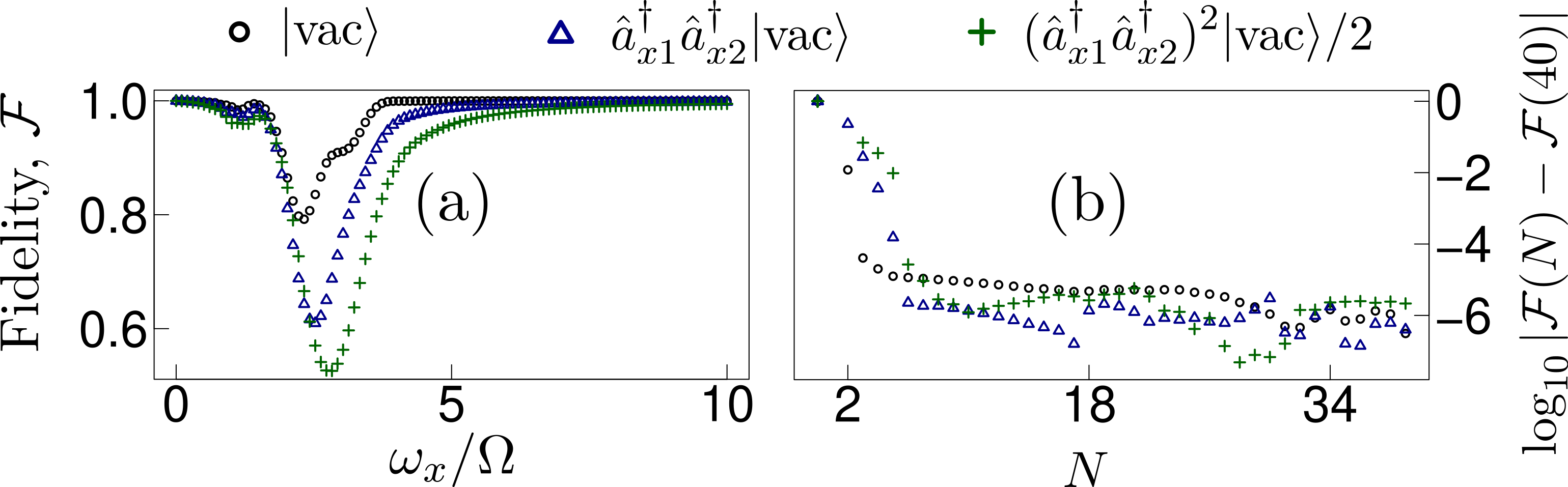}
\caption{(a) Fidelity of the $\text{C}_2(Z)$ gate as a function of the axial trap frequency $\omega_x$ when $\mathbb{J}=1.87\hbar\Omega\approx25.2\hbar/T$ at which the diabatic gate error is locally maximal as shown in Fig.~\ref{figure-uni-error}(b). The DDI between molecule a and molecule c(t) is given in Eq.~(\ref{unperturbedJ}), with $\ell_x/L_1$ equal to $0.12$, and the initial motional state for each data symbol is shown on the top. There is a trap-dipole resonance induced drop of the gate fidelity at around $\hbar\omega_x/\mathbb{J}=1$ as studied in Ref.~\cite{Lu2026gate2}, where the minimal of $\mathcal{F}$ differs for different initial motional states because of different ac Stark shifts~\cite{Lu2026gate,Lu2026gate2}. The number of harmonic modes is $N=20$ for both $\{\hat{a}_{x1},\hat{a}_{x1}^\dag\}$ and $\{\hat{a}_{x2},\hat{a}_{x2}^\dag\}$ by using QuTip~\cite{Johansson_2013}. (b) By setting $\omega_x=2.5\Omega$ which corresponds to the minimal $\mathcal{F}$ of the data shown with triangular symbols in (a), $ \log_{10}|\mathcal{F}(N)-\mathcal{F}(40)| $ drops quickly when $N$ increases from the minimally allowed value. Both the relative and absolute tolerances are $10^{-8}$ in the simulation. For faster simulation, we use $N=20$, the deviation of $\mathcal{F}$ from that of $N=40$ is several $10^{-6}$. Fidelity with motional states for the gate is derived in Appendix~\ref{app-fidelity}.
 \label{figure-linear} }
\end{figure}

\subsection{Reduction of DDI-coupled motional modes}\label{sec0qmo}
\subsubsection{Linear terms can be neglected}\label{sec05B1}
With Eqs.~(\ref{main-ac-03}) and~(\ref{main-phi-expansion2}) inserted in Eqs.~(\ref{Jact-main}), one can see that six quantized motional modes are coupled with DDI in the ancilla-assisted controlled-phase gate. If one directly uses a quantum mechanical treatment, the number of motional modes would be $N^6$ when truncating the number of motional quantum number at $N$, which challenges the required memory for the simulation. However, if $\omega_\alpha\gg \Omega,|\mathscr{J}_{\beta}/\hbar|$~\cite{picard_entanglement_2025} with $\alpha\in\{x,y,z\}$ and $\beta\in\{$ac, at, ct$\}$, the linear modes in Eq.~(\ref{main-phi-expansion2}) and those with coefficients $A_{\alpha,1}$ and $A_{\alpha,1}^\prime$ in Eq.~(\ref{Jact-main}) can be ignored. This is because transforming the Hamiltonians in Eq.~(\ref{Jact-main}) into a rotating frame with $e^{it\hat{H}_{\text{trap}}}(\hat{H}_{\text{trap}}+\sum_{\zeta} \mathscr{J}_\zeta)e^{-it\hat{H}_{\text{trap}}}-\hat{H}_{\text{trap}} $~\cite{Shi_2023} can yield a Hamiltonian where the creation and annihilation operators attain phase factors $e^{\pm i\omega_\xi t}$ with $\xi\in\{x,y,z\}$. These phase factors are fast rotating compared to other dynamics, and, hence, the terms they accompany can be ignored.

To verify the argument above, one can numerically study how weak the linear terms in Eq.~(\ref{Jact-main}) will be when he trap frequencies are large. Because of the different distances $2L_1$ and $L_1$ and the difference between $C_{\text{ct}}$ and $C_{\text{at(ac)}}$, the coefficients accompanying
the linear term in $\mathscr{J}_{\text{ct}}$ of Eq.~(\ref{Jact-main}) are 48 times smaller compared to that accompanying $\mathscr{J}_{\text{at(ac)}}$. Further, at $\vartheta=$arccos$(1/\sqrt{3})$, the coefficients $A_{\alpha,1}$ and $A_{\alpha,1}^\prime$ are comparable as shown in Tables~\ref{sensi-table} and~\ref{sensi-table2}. This means that when we test the argument that the first-order terms can be dropped, it is useful to omit $\mathscr{J}_{\text{ct}}$ since it is orders of magnitude smaller than $\mathscr{J}_{\text{at(ac)}}$. Therefore, one can test the correctness by including $\mathscr{J}_{\text{at(ac)}}$ in the DDI operator, and we can take the following DDI as from Eq.~(\ref{Jact-main}),
\begin{eqnarray}
\mathscr{J}_{\text{ac}} &\leftarrow &  \frac{C_{\text{ac}} }{L_1^3}  \left[ \sin^2\vartheta \left(  -1 - \frac{2i}{L_1\sin\vartheta} x_{\text{ac}}     \right) -\frac{   A_{y,1} y_{\text{ac}}  + A_{z,1} z_{\text{ac}}   } {L_1}    \right]\nonumber\\
&=&\frac{C_{\text{ac}} }{L_1^3}  \left[ - \frac{2}{3} - \frac{2\sqrt{
6}i}{3}\frac{ x_{\text{ac}}}{L_1 }  + \frac{4\sqrt{
6}}{9} \frac{    y_{\text{ac}}    } {L_1}   + \frac{10\sqrt{
3}}{9}  \frac{   z_{\text{ac}}   } {L_1}   \right] ,\nonumber\\
\mathscr{J}_{\text{at}} &\leftarrow   &  \frac{C_{\text{at}} }{L_1^3}  \left[ \sin^2\vartheta\left(  -1 - \frac{2i}{L_1\sin\vartheta} x_{\text{at}}     \right) + \frac{   A_{y,1} y_{\text{at}}  + A_{z,1} z_{\text{at}}   } {L_1}    \right] \nonumber\\
&=&\frac{C_{\text{at}} }{L_1^3}  \left[ - \frac{2}{3} - \frac{2\sqrt{
6}i}{3}\frac{ x_{\text{at}}}{L_1 }  + \frac{4\sqrt{
6}}{9} \frac{    y_{\text{at}}    } {L_1}   + \frac{10\sqrt{
3}}{9}  \frac{   z_{\text{at}}   } {L_1}   \right] ,\label{Jact-main1}
\end{eqnarray}
where the value of $\vartheta=$arccos$(1/\sqrt{3})$ is inserted. To compare the strengths of the three terms with creation and annihilation operators in $\mathscr{J}_{\text{ac}}$, we note that $\xi_{\text{ac}}$ have similar forms with $\xi\in\{x,y,z\}$, but possess different $\ell_\xi$. Then, the coupling strengths have the following ratio,
\begin{eqnarray}
  \left| - \frac{2\sqrt{6}i}{3}\frac{\ell_x}{L_1 } \right| :   \left| \frac{4\sqrt{
6}}{9} \frac{   \ell_y   } {L_1}  \right| :   \left|\frac{10\sqrt{
3}}{9}  \frac{   \ell_z  } {L_1}\right| =1: \frac{2}{3} \left(\frac{\omega_x}{\omega_y}\right)^{\frac{1}{2}}:  \frac{5}{3\sqrt{2}} \left(\frac{\omega_x}{\omega_z}\right)^{\frac{1}{2}}, \label{ratio1term}
\end{eqnarray}
which also applies to the terms in $\mathscr{J}_{\text{at}}$. If we consider usual trap geometries where the two radial directions have the same trap frequencies, which are much larger than that along the axial direction, the dominant term would be the one along the axial direction. For example, if we take the geometry in Ref.~\cite{ruttley_long-lived_2025}, then $(\omega_x,\omega_{y,z})=2\pi\times(0.4,3,3)$~kHz, so that the ratio in Eq.~(\ref{ratio1term}) is about $1:0.24:0.43$. If we remove the latter two terms, we should increase the strength of the first term by a factor of $\sqrt{1+0.24^2+0.43^2}\approx1.1$ to more reasonably reflect the coupling between DDI and QM. In other words, we use $  \frac{C_{\text{ac}} }{L_1^3}  ( - \frac{2}{3} -1.1\cdot \frac{2\sqrt{
6}i}{3}\frac{ x_{\text{ac}}}{L_1 })$ as $\mathscr{J}_{\text{ac}}$ and $  \frac{C_{\text{at}} }{L_1^3}  ( - \frac{2}{3} -1.1\cdot \frac{2\sqrt{
6}i}{3}\frac{ x_{\text{at}}}{L_1 })$ as $\mathscr{J}_{\text{at}}$, which can be written as
\begin{eqnarray}
\mathscr{J}_{\text{ac(at)}}&=& -\mathbb{J} \left\{1+1.1\cdot\sqrt{6}i \frac{\ell_x}{L_1}\left[ \frac{\sqrt{3}}{2}  (\hat{a}_{x1}^\dag +  \hat{a}_{x1}) \mp \frac{1}{2}(\hat{a}_{x2}^\dag + \hat{a}_{x2}) \right] \right\},\label{unperturbedJ}
\end{eqnarray}
where
\begin{eqnarray}
\mathbb{J} &=& \frac{1}{4\pi\epsilon_0}\frac{\text{\textdong}^2}{3L_1^3}, ~~ 
\end{eqnarray}
with the help of the definition in Eq.~(\ref{main-ac-03}) and $C_{\text{ac}} =C_{\text{at}} =\frac{1}{4\pi\epsilon_0}\frac{\text{\textdong}^2}{2}$ as from Appendix~\ref{app-ddi},
where the minus~(plus) sign in the bracket of Eq.~(\ref{unperturbedJ}) is for $\mathscr{J}_{\text{ac(at)}}$.
By using the interaction in Eq.~(\ref{unperturbedJ}), Fig~\ref{figure-linear}(a) shows the fidelity of the ancilla-assisted $\text{C}_2(Z)$ gate with three types of initial motional states. The round, triangular, and cross symbols are from states with zero, one, and two motional quanta in both the $\{\hat{a}_{x1},\hat{a}_{x1}^\dag\}$ and $\{\hat{a}_{x2},\hat{a}_{x2}^\dag\}$ modes. Due to the slow simulation with large dimension when using QuTip~\cite{Johansson_2013}, we use Fock states in the motional modes instead of thermal states. As shown in Refs.~\cite{Lu2026gate,Lu2026gate2}, using thermal states can yield results similar to that with Fock states, and the gate fidelity can even be higher with thermal states~\cite{Lu2026gate}. Figure~\ref{figure-linear}(a) shows that there is a trap-dipole resonance for the gate, but when $\hbar\omega_x$ is much smaller, or much larger than $\mathbb{J}$, the fidelity shoots up to near 100$\%$, indicating that the perturbing term linear to $\ell_x/L_1$ in Eq.~(\ref{main-ac-03}) has negligible effect when $\hbar\omega_x$ is large compared to DDI.

The results in Fig~(\ref{figure-linear}) are obtained by truncating the motional states at $N=20$. To show that this truncation does not bring large error, we study the relative error with different numbers of motional states used in the simulation. Figure~\ref{figure-linear}(b) shows that increasing $N$ from 20 to some larger number barely changes the results. Because the accuracy used in the simulation is $10^{-8}$, the relative error shown in Fig~\ref{figure-linear}(b) is on the level of $10^{-6}$ when $N=20$. This means that the results in Fig~\ref{figure-linear}(a) have relative errors less than $10^{-5}$, and it indicates that the conclusion drawn from it is reliable on the level of $10^{-5}$.

\subsubsection{Anisotropic trapping limit}
The number of second-order terms can also be reduced in the limit of anisotropic trapping. We consider the example of Ref.~\cite{ruttley_long-lived_2025} where $(\omega_x,\omega_{y,z})=2\pi\times(0.4,3,3)$~kHz so that the value of $\omega_x/\omega_{y(z)}$ is small. In other words, $\ell_x$ is much larger than $\ell_{y(z)}$.

By preserving the zeroth and the second-order terms in the DDI operators, we have
\begin{eqnarray}
\mathscr{J}_{\text{ac}} &\approx & - \frac{C_{\text{ac}} }{L_1^3}  \left[ \sin^2\vartheta\left(  1-\frac{2}{(L_1\sin\vartheta)^2}x_{\text{ac}}^2- \frac{2i}{(L_1\sin\vartheta)^2} x_{\text{ac}} y_{\text{ac}}     \right)  +\frac{2i}{\sin\vartheta}\frac{ x_{\text{ac}}}{L_1 }   \frac{   A_{y,1} y_{\text{ac}}  + A_{z,1} z_{\text{ac}}   } {L_1} \right.\nonumber\\
&&\left. + \frac{A_{xx,2} x_{\text{ac}}^2  + A_{yy,2}y_{\text{ac}}^2+ A_{zz,2} z_{\text{ac}}^2 + A_{yz,2} y_{\text{ac}}z_{\text{ac}}}{L_1^2} \right] ,\nonumber\\
\mathscr{J}_{\text{at}} &\approx & - \frac{C_{\text{at}} }{L_1^3}  \left[ \sin^2\vartheta\left(  1-\frac{2}{(L_1\sin\vartheta)^2}x_{\text{at}}^2- \frac{2i}{(L_1\sin\vartheta)^2} x_{\text{at}} y_{\text{at}}     \right)+ \frac{2i}{\sin\vartheta}\frac{ x_{\text{at}}}{L_1 } \frac{   A_{y,1} y_{\text{at}}  + A_{z,1} z_{\text{at}}   } {L_1}\right.\nonumber\\
&&\left.   + \frac{A_{xx,2} x_{\text{at}}^2  + A_{yy,2}y_{\text{at}}^2+ A_{zz,2} z_{\text{at}}^2 + A_{yz,2} y_{\text{at}}z_{\text{at}}}{L_1^2} \right] ,\nonumber\\
\mathscr{J}_{\text{ct}} &\approx & \frac{C_{\text{ct}}}{(2L_1)^3}  \frac{A_{xx,2}^\prime  x_{\text{ct}}^2  + A_{yy,2}^\prime y_{\text{ct}}^2+ A_{zz,2}^\prime  z_{\text{ct}}^2 + A_{yz,2}^\prime  y_{\text{ct}}z_{\text{ct}}}{(2L_1)^2} . \label{Jact-main2ed}
\end{eqnarray}
If $\omega_x/\omega_{y(z)}=2/15$~\cite{ruttley_long-lived_2025}, then the coupling strengths between the bosonic operators derived from $x_\alpha^2,y_\alpha^2,z_\alpha^2,x_\alpha y_\alpha,x_\alpha z_\alpha,y_\alpha z_\alpha$ in $\mathscr{J}_\alpha$ is
\begin{eqnarray}
 && |A_{xx,2}-2|\ell_x^2 :   |A_{yy,2} |\ell_y^2 :  |A_{zz,2} |\ell_z^2  : \left| \frac{2i}{\sin\vartheta}A_{y,1}-2i \right| \ell_x\ell_y: \left|\frac{2i}{\sin\vartheta}A_{z,1} \right|\ell_x\ell_z : |A_{yz,2}|\ell_y\ell_z\nonumber\\
 &\approx& 1: 0.022: 0.111: 0.639: 0.645: 0.314,
\end{eqnarray}
for both $\alpha=$ac and at, which is estimated by the data in Table~\ref{sensi-table}. One can see that the quadratic terms are dominated by the one along $\mathbf{x}$. To capture the leading effect of these terms, we can drop off the two terms $y_\alpha^2$ and $z_\alpha^2$ with both $\alpha=$ac and at. Among the three cross terms, the two with $x_\alpha y_\alpha$ and $x_\alpha z_\alpha$ have a strength comparable to
that with $x_\alpha^2$. However, as discussed in Sec.~\ref{sec05B1}, if we transform the Hamiltonians in Eq.~(\ref{Jact-main}) into a rotating frame with $e^{it\hat{H}_{\text{trap}}}(\hat{H}_{\text{trap}}+\sum_{\zeta} \mathscr{J}_\zeta)e^{-it\hat{H}_{\text{trap}}}-\hat{H}_{\text{trap}} $, we can have a Hamiltonian where the creation and annihilation operators attain phase factors $e^{\pm i\omega_\xi t}$ with $\xi\in\{x,y,z\}$. For the cross terms $x_\alpha y_\alpha$ and $x_\alpha z_\alpha$, the $x_\alpha$ part attain a phase factors $e^{\pm i\omega_x t}$, while the $y_\alpha$ or $z_\alpha$ part get $e^{\pm i\omega_{y(z)} t}$, so that the final phase factors would be $e^{\pm i(\omega_x\pm\omega_{y(z)}) t}$. Due to $\omega_x/\omega_{y(z)}\ll1$, the resulting phase factors lead to fast rotating dynamics as in the linear terms studied in Sec.~\ref{sec05B1}. As a consequence, the terms $x_\alpha y_\alpha$ and $x_\alpha z_\alpha$ can be ignored.

In the remaining two terms $x_\alpha^2$ and $y_\alpha z_\alpha$ in $\mathscr{J}_\alpha$, since there is no coupling between them, the error from them is uncorrelated. In this case, one can estimate the error by taking the leading term, i.e., $x_\alpha^2$, but increasing its magnitude by a factor of $\sqrt{1+0.314^2}\approx 1.05$, so that
\begin{eqnarray}
\mathscr{J}_{\text{ac(at)}}&=& -\mathbb{J} \left\{1+1.05\cdot \frac{3(A_{xx,2}-2)}{2} \frac{\ell_x^2}{L_1^2}\left[ \frac{\sqrt{3}}{2}  (\hat{a}_{x1}^\dag + \hat{a}_{x1}) \mp \frac{1}{2} (\hat{a}_{x2}^\dag + \hat{a}_{x2}) \right]^2 \right\},\label{unperturbedJ2}
\end{eqnarray}
where $A_{xx,2}=-2/3$ from Table~\ref{sensi-table}.

\begin{figure}
\includegraphics[width=5.0in]
{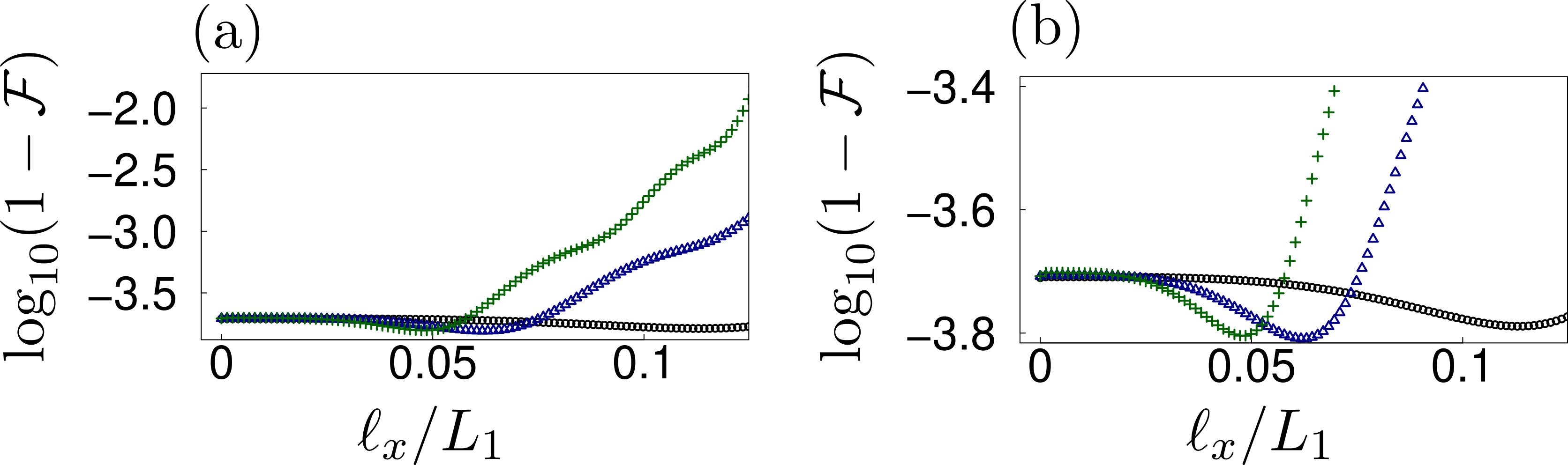}
\caption{Fidelity~(in log scale) of the $\text{C}_2(Z)$ gate as a function of $\ell_x/L_1$ when $\mathbb{J}=1.87\hbar\Omega$ and $\omega_x=10\Omega$. The DDI between molecule a and molecule c(t) is given in Eq.~(\ref{unperturbedJ2})+Eq.~(\ref{Jact-main3ed})+the linear term in Eq.~(\ref{unperturbedJ}). As in Fig.~\ref{figure-linear}, the initial motional state is $\lvert \text{vac}\rangle$, $\hat{a}_{x1}^\dag \hat{a}_{x2}^\dag \lvert \text{vac}\rangle$, and $(\hat{a}_{x1}^\dag \hat{a}_{x2}^\dag )^2 \lvert \text{vac}\rangle/2$ for the data with circular, triangular, and cross symbols, respectively. Each motional mode has up to $N=20$ motional states. (a,b) show the same data, where (b) is the close-up around the maximal $\mathcal{F}$. (b) shows that the gate fidelity first increases when $\ell_x$ increases. This is because when $\ell_x$ increases, the term $(\hat{a}_{x1}^\dag + \hat{a}_{x1})^2$ and $(\hat{a}_{x2}^\dag + \hat{a}_{x2} )^2$ in Eq.~(\ref{unperturbedJ2}) will yield constants, which effectively change $\mathscr{J}_{\text{ac(at)}}$ away from the value $1.87\hbar\Omega$. Because $\mathscr{J}_{\text{ac(at)}}=1.87\hbar\Omega$ corresponds to a local minimum of $\mathcal{F}$ as shown Fig.~\ref{figure-uni-error}(b), the gate fidelity will increase when $\ell_x$ increases within a certain range.
 \label{figure-second} }
\end{figure}

The coupling strengths for $x_{\text{ct}}^2, y_{\text{ct}}^2, z_{\text{ct}}^2$, and $y_{\text{ct}}z_{\text{ct}}$ in $\mathscr{J}_{\text{ct}}$, as shown in Eq.~(\ref{Jact-main2ed}), have a ratio of $1: 0.756: 0.622:0.629$ according to Table~\ref{sensi-table2} with $\omega_x/\omega_{y(z)}=2/15$~\cite{ruttley_long-lived_2025}, and, hence, each term should not be ignored. If we take all these modes and truncate the motional modes with up to $N$ motional quanta as in Fig.~\ref{figure-linear}, the number of motional states is $N^6$, which is over $6\times10^7$ if $N=20$, rendering extreme difficulty in a numerical study. To study the gate error with available resources, we note that though the terms with $y_{\text{ct}}^2, z_{\text{ct}}^2$, and $y_{\text{ct}}z_{\text{ct}}$ are coupled together in $\mathscr{J}_{\text{ct}}$, they are uncoupled with the term $x_{\text{ct}}^2$. Therefore, we can estimate the impact of $\mathscr{J}_{\text{ct}}$ by increasing the coupling strength with $x_{\text{ct}}^2$ by a factor of $(1+ 0.756^2+ 0.622^2+0.629^2)^{1/2}\approx1.53$ when dropping off the terms with $y_{\text{ct}}^2, z_{\text{ct}}^2$, and $y_{\text{ct}}z_{\text{ct}}$. This means that we can use
\begin{eqnarray}
\mathscr{J}_{\text{ct}} &\approx & -1.53\cdot \mathbb{J}  \frac{A_{xx,2}^\prime}{32}  \frac{ \ell_x^2}  {L_1^2}  (\hat{a}_{x2}^\dag + \hat{a}_{x2})^2,\label{Jact-main3ed}
\end{eqnarray}
where $A_{xx,2}^\prime=1$ from Table~\ref{sensi-table2} and we have used the relation $C_{\text{ct}}=-C_{\text{ac}}/3$. Then, because the term with $x_{\text{ct}}^2$ involves motional modes that are included in Eq.~(\ref{unperturbedJ2}), the total number of motional states is still $N^2$ when including Eq.~(\ref{Jact-main3ed}) and $\mathscr{J}_{\text{ct}}$.

By including the DDI given in Eq.~(\ref{unperturbedJ2}) and Eq.~(\ref{Jact-main3ed}), and also the linear term proportional to $\ell_x/L_1$ in Eq.~(\ref{unperturbedJ}), we show the gate fidelity~(in log scale) of the $\text{C}_2(Z)$ gate as a function of $\ell_x/L_1$ in Fig.~\ref{figure-second}. As in Fig.~\ref{figure-linear}, here we use $\mathbb{J}=1.87\hbar\Omega$ where the gate fidelity is locally minimal as shown in Fig.~\ref{figure-uni-error}(b). To avoid trap-dipole resonance~\cite{Lu2026gate2}, we set $\omega_x=10\Omega$. The three types of symbols correspond to the same meaning in Fig.~\ref{figure-linear}, i.e., the initial motional state is $\lvert \text{vac}\rangle$, $\hat{a}_{x1}^\dag \hat{a}_{x2}^\dag \lvert \text{vac}\rangle$, and $(\hat{a}_{x1}^\dag \hat{a}_{x2}^\dag )^2 \lvert \text{vac}\rangle/2$ for the data with circular, triangular, and cross symbols, respectively. Here, each of the two motional modes $\{\hat{a}_{x1}^\dag, \hat{a}_{x1}\}$ and $\{\hat{a}_{x2}^\dag, \hat{a}_{x2}\}$ has up to $N=20$ states.

A surprising result in Fig.~\ref{figure-second} is that when $\ell_x$ increases, the gate fidelity can even increase compared to the case when no QM exists. To understand this, one can expand Eq.~(\ref{unperturbedJ2}),
\begin{eqnarray}
\mathscr{J}_{\text{ac(at)}}&=& -\mathbb{J} \left\{1+1.05\cdot \frac{3(A_{xx,2}-2)}{2} \frac{\ell_x^2}{L_1^2}\left[1+  \frac{3}{4} \left( 2\hat{a}_{x1}^\dag \hat{a}_{x1}+ (\hat{a}_{x1}^\dag)^2 + \hat{a}_{x1}^2 \right)\right.\right.\nonumber\\
&& \left.\left. +  \frac{1}{4} \left( 2\hat{a}_{x2}^\dag \hat{a}_{x2}+ (\hat{a}_{x2}^\dag)^2 + \hat{a}_{x2}^2 \right)  \right] \right\},\label{unperturbedJ3}
\end{eqnarray}
where the terms $(\hat{a}_{x1(2)}^\dag)^2$ and $\hat{a}_{x1(2)}^2$ induce fast rotating dynamics as studied in Sec.~\ref{sec05B1}. The remaining term $1+ \frac{3}{2}\hat{a}_{x1}^\dag \hat{a}_{x1} +  \frac{1}{2} \hat{a}_{x2}^\dag \hat{a}_{x2}$ is equal to $1, 3,$ and 5 for the initial motional states $\lvert \text{vac}\rangle$, $\hat{a}_{x1}^\dag \hat{a}_{x2}^\dag \lvert \text{vac}\rangle$, and $(\hat{a}_{x1}^\dag \hat{a}_{x2}^\dag )^2 \lvert \text{vac}\rangle/2$, respectively. As shown in Fig.~\ref{figure-uni-error}(b), when the DDI deviates from the value corresponding to the locally minimal $\mathcal{F}$, the fidelity will increase. But when the deviation exceeds a certain value, the increase of the fidelity stops and turns to decrease. Because for the states with more motional quanta, the deviation is faster, $\mathcal{F}$ begins to decrease at a smaller $\ell_x$ as shown in Fig.~\ref{figure-second}(b). This means that when a gate is designed with parameters at the locally minimal $\mathcal{F}$, the gate can be robust against QM of the molecules.

\begin{table}[ht]
  \centering
  \caption{Logarithm~($\log_{10}$) of the average infidelity of the ancilla-assisted $\text{C}_2(Z)$ gate by sampling $\mathbb{J}$ over $(0.75,~1.25)\cdot 1.87\hbar\Omega$. Other parameters are the same as those in Fig.~\ref{figure-second}. The gate error is from the diabatic, QM, and fluctuation of $\mathbb{J}$, where the last can arise from shot-to-shot fluctuation of the position of trap centers.  \label{table-shot}  }
  \begin{tabular}{|c|c|c|c|}
    \hline
\diagbox{Motional state}{$\ell/L_1$} & $0.04$ & $0.08$ &  $0.12$ \\
     \hline
$\lvert \text{vac}\rangle$ &   -3.87    &  -3.80  & -3.63  \\
\hline$\hat{a}_1^\dag\hat{a}_2^\dag\lvert \text{vac}\rangle$  &-3.83
&-3.53
&-2.85 \\
\hline$(\hat{a}_1^\dag\hat{a}_2^\dag)^2\lvert \text{vac}\rangle/2$  & -3.74   &  -3.17    & -2.08 \\
   \hline
  \end{tabular}

  \end{table}

As shown below Eq.~(\ref{omega-impure}), DDI between the data molecule and the ancilla equal to $h\times1$~kHz corresponds to $|OD_1|=|OD_2|\approx1.29~\mu$m with the molecular configuration of Fig.~\ref{figure-1d}. If $\ell_x=0.1~\mu$m~\cite{picard_entanglement_2025}, leading to $\ell_x/|OD_1|\approx0.078$, the gate error shown in Fig.~\ref{figure-uni-error} is $6.089\times10^{-4}$ when the initial state is $(\hat{a}_{x1}^\dag \hat{a}_{x2}^\dag )^2 \lvert \text{vac}\rangle/2$, which means that after deducting the diabatic error, the extra error from the QM is $4.13\times10^{-4}$.

Another possible error is that the center of the traps can fluctuate, which will effectively change the value of $\mathbb{J}$. With a $1/4$ change around the desired value of $\mathbb{J}$, Table~\ref{table-shot} shows the averaged gate error from the diabatic error and the QM of the molecules inside the traps. At $\ell_x/|OD_1|\approx0.078$, the average fidelity is very near to the one shown in Fig.~\ref{figure-uni-error}, which means that a $25\%$ variation of $\mathbb{J}$ barely alters the gate accuracy, and, hence, the result in Table~\ref{table-shot} is not included in Table~\ref{table-fid}. One can see that the error with the same $\ell_x/L_1$ is larger compared to that in Ref.~\cite{Lu2026gate}, which is because the molecular configuration here leads to stronger fluctuation as compared to the two-molecule system of Ref.~\cite{Lu2026gate}. However, the ancilla-assisted gate does not need the single-site addressing as required in Ref~\cite{Lu2026gate}, rendering a more experimentally friendly protocol.

\begin{figure}
\includegraphics[width=5.5in]
{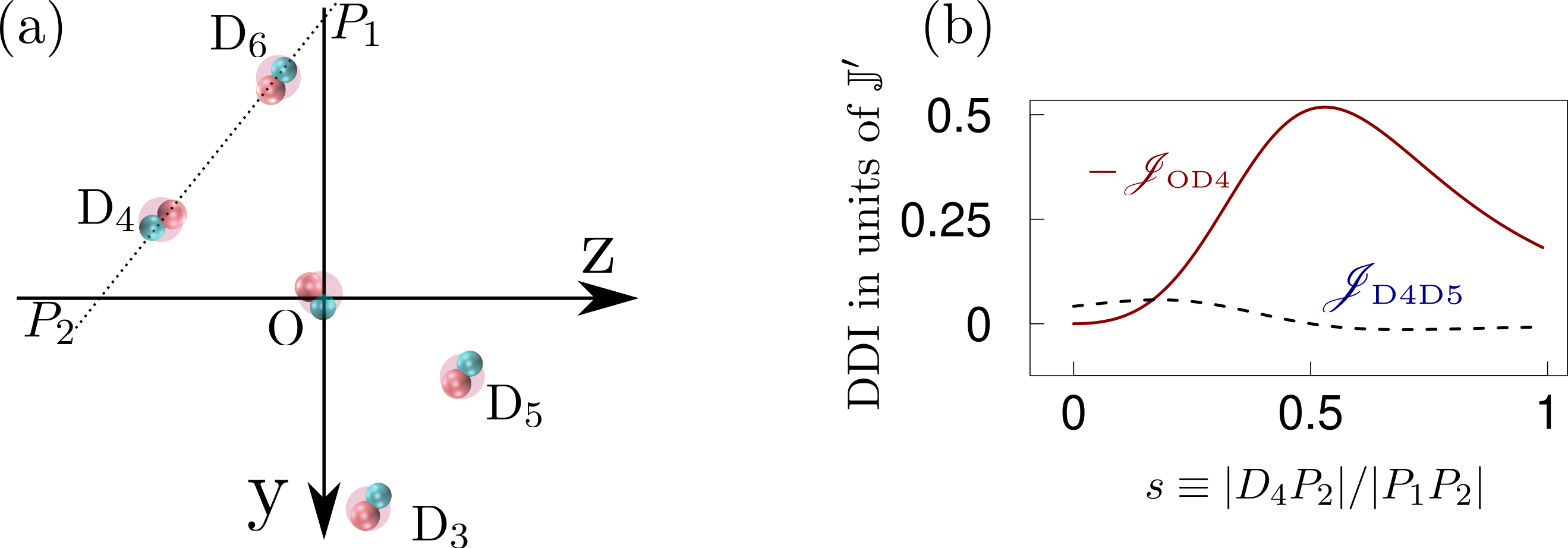}
\caption{(a) Configuration of molecular qubits for realizing a three-qubit or four-qubit controlled phase gates, as in Fig.~\ref{figure-1d}. The four data molecules are at $D_{3},D_{5}, D_4$ and $D_{6}$. For the three-qubit $\text{C}_3(Z)$ gate, we can use, e.g., the three data molecules at $D_{3},D_{4},$ and $D_{6}$. The three sites $D_{j},D_{j+2},D_{j+4}$ are on one line for both $j=1$ and $j=2$, tilted from $\mathbf{z}$ by an angle $\vartheta$. The two sites $D_4$ and $D_5$ are symmetric to each other with respect to $O$; similar for the two sites $D_3$ and $D_6$. (b) The solid curve shows the magnitude of DDI between the molecules at $D_4$ and $O$, as calculated in Eq.~(\ref{4qubit-01}); the dashed curve shows the DDI between the data molecules at $D_4$ and $D_5$, as calculated in Eq.~(\ref{4qubit-02}). Here, $\mathbb{J}^\prime$ is defined as $\frac{1}{4\pi\epsilon_0}\frac{\text{\textdong}^2}{L_2^3}$, with $L_2$ being the distance between $O$ and $P_2$.  \label{figure-6m} }
\end{figure}

\section{Multi-qubit controlled phase gates}\label{sec06}
\subsection{Gate mechanism}\label{sec06a}
The dark-state mechanism can be extended to a multi-qubit system. Because the central requirement for the dark-state condition is that no DDI occurs between any data molecules, the physical picture for an $\mathbb{N}$-qubit dark-state molecular gate is the same to the case of $\mathbb{N}=2$. Therefore, we can use the cases of $\mathbb{N}=3,4$ to show the essence of a multi-qubit controlled-phase gate with an arbitrary controlled phase. In particular, we consider data molecules trapped at $D_{3},D_{5}, D_4$ and $D_{6}$ of Fig.~\ref{figure-6m}(a). The sites $D_{j},D_{j+2},D_{j+4}$ are on one line for both $j=1$ and $j=2$, tilted from $\mathbf{z}$ by an angle $\vartheta$. There is no DDI between the two data molecules at $D_{3},D_{5}$; similarly for the molecules at $D_4$ and $D_{6}$. The DDI between one at $\{D_3,D_5\}$ and another at $\{D_4,D_6\}$ can be weak when their distances are large. Then, we can approximately have negligible DDI between any two data molecules, and the dark-state mechanism can proceed.

\subsubsection{Three-qubit gate}\label{sec06a1}
 There are four different classes of input states:\newline (i) $ \lvert \downarrow_{\text{a}},\downarrow,\downarrow,\downarrow\rangle$;\newline (ii) $ \lvert \downarrow_{\text{a}},\uparrow,\downarrow,\downarrow\rangle$, $ \lvert \downarrow_{\text{a}},\downarrow,\uparrow,\downarrow\rangle$, $ \lvert \downarrow_{\text{a}},\downarrow,\downarrow,\uparrow\rangle$; \newline (iii) $ \lvert \downarrow_{\text{a}},\uparrow,\uparrow,\downarrow\rangle$, $ \lvert \downarrow_{\text{a}},\uparrow,\downarrow,\uparrow\rangle$, $ \lvert \downarrow_{\text{a}},\downarrow,\uparrow,\uparrow\rangle$;\newline (iv) $ \lvert \downarrow_{\text{a}},\uparrow,\uparrow,\uparrow\rangle$,\newline where the three state symbols without subscript denote the states of the molecules at $D_3,D_4,D_6$, respectively.

By using the two-pulse protocol in Sec.~\ref{sec02}, the input state in class (i),
$\lvert  \downarrow_{\text{a}},\downarrow,\downarrow,\downarrow\rangle$, evolves to $-i\lvert  e_{\text{a}},\downarrow,\downarrow,\downarrow\rangle$ after the first pulse. In the second pulse with a phase shift $\pi-\varphi$, i.e., with the Rabi frequency changed to $-e^{-i\varphi}\Omega_\mu$, the input state of class (i) evolves to $e^{i\varphi}\lvert  \downarrow_{\text{a}},\downarrow,\downarrow,\downarrow\rangle$. So the overall effect is a phase change $\varphi$.

The three states in class (ii) have one data molecule initialised in $\uparrow$, so when the microwave pulse is sent, there will be a Hamiltonian similar to Eq.~(\ref{adia-01}). Then, a dark-state evolution will appear for the states in class (ii), resulting in no change to the state.

The three states in class (iii) have two data molecules initialised in $\uparrow$. When the microwave pulse is sent, there will be a Hamiltonian similar to Eq.~(\ref{adia2-01}), which ensures a dark-state evolution. Therefore the states in class (iii) will also stay intact.

For the state in class (iv), there is no similar Hamiltonian for describing its dynamics in Sec.~\ref{sec02}. To analyze it, we note that when there is no DDI between the data molecules, the Hamiltonian for it is
\begin{eqnarray}
\hat{H}_{\text{iv}} &=&\left(
\begin{array}{ccc}
0&A_3& 0\\
 A_3&0 &\hbar \frac{\Omega_\mu^\ast}{2} \\
 0 &\hbar \frac{\Omega_\mu}{2} & 0
\end{array}
\right)\label{sec06-eq1}
\end{eqnarray}
in the basis of $ \{    \lvert \Lambda_3 \rangle      ,  \lvert e_{\text{a}},\uparrow,\uparrow,\uparrow\rangle     ,  \lvert \downarrow_{\text{a}},\uparrow,\uparrow,\uparrow\rangle  \} $, where $A_3=\sqrt{\mathscr{J}_{\text{\tiny OD3}}^2+\mathscr{J}_{\text{\tiny OD4}}^2+\mathscr{J}_{\text{\tiny OD6}}^2 }$, with $\mathscr{J}_{\text{\tiny ODj}}$ the DDI between the ancilla at $O$ and the data molecule at $D_j$, and
\begin{eqnarray}
 \lvert \Lambda_3 \rangle  &=& \frac{1}{A_3}\left( \mathscr{J}_{\text{\tiny OD3}}\lvert \uparrow_{\text{a}},e,\uparrow,\uparrow\rangle +\mathscr{J}_{\text{\tiny OD4}}\lvert \uparrow_{\text{a}},\uparrow,e,\uparrow\rangle+\mathscr{J}_{\text{\tiny OD6}}\lvert \uparrow_{\text{a}},\uparrow,\uparrow,e\rangle  \right).\label{Lambda-def}
\end{eqnarray}
where the subscript 3 is used because a 3-qubit gate is studied here. One can see that the above Hamiltonian can support the dark-state evolution outlined in Sec.~\ref{sec02}. Therefore, the states in classes (ii, iii) and (iv) all stay intact, and the only state that changes is the one in class (i). Because the state of the ancilla is restored, we have the $\text{C}_3(Z)$ gate that induces
\begin{eqnarray}
\lvert \downarrow_{\text{a}},\downarrow,\downarrow,\downarrow\rangle&\mapsto& e^{i\varphi}\lvert \downarrow_{\text{a}}, \downarrow,\downarrow,\downarrow\rangle,\label{adia-ccz}
\end{eqnarray}
which occurs only when the input state is $\lvert \downarrow_{\text{a}}, \downarrow,\downarrow,\downarrow\rangle$.

The above consideration assumes perfectly annulled DDI between any data molecules. In practice, QM of the molecules can bring up nonzero DDI between data molecules. Moreover, for the configuration in Fig.~\ref{figure-6m}(a), there will be nonzero DDI between the molecule at $D_3$ and the molecule at $D_{4(6)}$. For the general case, we consider that DDI can arise between any two data molecules, so that Eq.~(\ref{sec06-eq1}) should be
\begin{eqnarray}
\hat{H}_{\text{iv}} &=&\left(
\begin{array}{ccccc}
0&\mathscr{J}_{\text{\tiny D4D3}}&\mathscr{J}_{\text{\tiny D6D3}}&\mathscr{J}_{\text{\tiny OD3}} & 0\\
\mathscr{J}_{\text{\tiny D4D3}}^\ast&0&\mathscr{J}_{\text{\tiny D6D4}}&\mathscr{J}_{\text{\tiny OD4}}& 0\\
\mathscr{J}_{\text{\tiny D6D3}}^\ast&\mathscr{J}_{\text{\tiny D6D4}}^\ast&0&\mathscr{J}_{\text{\tiny OD6}}& 0\\
 \mathscr{J}_{\text{\tiny OD3}}^\ast & \mathscr{J}_{\text{\tiny OD4}}^\ast& \mathscr{J}_{\text{\tiny OD6}}^\ast&0 &\hbar \frac{\Omega_\mu^\ast}{2} \\
 0&0&0 &\hbar \frac{\Omega_\mu}{2} & 0
\end{array}
\right),\label{sec06-eq2}
\end{eqnarray}
in the basis of $ \{ \lvert \uparrow_{\text{a}},e,\uparrow,\uparrow\rangle,\lvert \uparrow_{\text{a}},\uparrow,e,\uparrow\rangle,\lvert \uparrow_{\text{a}},\uparrow,\uparrow,e\rangle    \lvert   ,  \lvert e_{\text{a}},\uparrow,\uparrow,\uparrow\rangle     ,  \lvert \downarrow_{\text{a}},\uparrow,\uparrow,\uparrow\rangle  \} $, where we include Hermitian conjugate to all the DDI operators for the sake of clarity, but in practice the DDI between any two data molecules are real as shown in Appendix~\ref{app-ddi}.

\subsubsection{Four-qubit gate}\label{sec06a2}
Here, we extend the study in Sec.~\ref{sec06a1} and consider a four-qubit gate, with the four molecules trapped at $D_3,D_4,D_5,D_6$ in Fig.~\ref{figure-6m} as an example. There are five different classes of input states:\newline (i) $ \lvert \downarrow_{\text{a}},\downarrow,\downarrow,\downarrow,\downarrow\rangle$;\newline (ii) $ \lvert \downarrow_{\text{a}},\uparrow,\downarrow,\downarrow,\downarrow\rangle$, $ \lvert \downarrow_{\text{a}},\downarrow,\uparrow,\downarrow,\downarrow\rangle$, $ \lvert \downarrow_{\text{a}},\downarrow,\downarrow,\uparrow,\downarrow\rangle$,$ \lvert \downarrow_{\text{a}},\downarrow,\downarrow,\downarrow,\uparrow\rangle$; \newline (iii) $ \lvert \downarrow_{\text{a}},\uparrow,\uparrow,\downarrow,\downarrow\rangle$, $ \lvert \downarrow_{\text{a}},\uparrow,\downarrow,\uparrow,\downarrow\rangle$, $ \lvert \downarrow_{\text{a}},\uparrow,\downarrow,\downarrow,\uparrow\rangle$, $ \lvert \downarrow_{\text{a}},\downarrow,\uparrow,\uparrow,\downarrow\rangle$, $ \lvert \downarrow_{\text{a}},\downarrow,\uparrow,\downarrow,\uparrow\rangle$, $ \lvert \downarrow_{\text{a}},\downarrow,\downarrow,\uparrow,\uparrow\rangle$;\newline (iv) $ \lvert \downarrow_{\text{a}},\uparrow,\uparrow,\uparrow,\downarrow\rangle$,$ \lvert \downarrow_{\text{a}},\uparrow,\uparrow,\downarrow,\uparrow\rangle$,$ \lvert \downarrow_{\text{a}},\uparrow,\downarrow,\uparrow,\uparrow\rangle$,$ \lvert \downarrow_{\text{a}},\downarrow,\uparrow,\uparrow,\uparrow\rangle$;\newline (v) $ \lvert \downarrow_{\text{a}},\uparrow,\uparrow,\uparrow,\uparrow\rangle$,\newline
where the four state symbols without subscript denote the states of the molecules at $D_3,D_4,D_5,D_6$, respectively. Since the state $\downarrow$ is not microwave excited, nor coupled by DDI, it does not change. So, one can see that the state dynamics for the states in (ii-iv) are exactly the same to those for the three class of states (ii-iv) listed in Sec.~\ref{sec06a1}. The only extra class of states in the four-qubit gate is the state in class (v). As in Eq.~(\ref{sec06-eq2}), when we consider that DDI can arise between any two data molecules, so that the Hamiltonian for the state in class (v) is given by
\begin{eqnarray}
\hat{H}_{\text{v}} &=&\left(
\begin{array}{cccccc}
0&\mathscr{J}_{\text{\tiny D4D3}}&\mathscr{J}_{\text{\tiny D5D3}}&\mathscr{J}_{\text{\tiny D6D3}}&\mathscr{J}_{\text{\tiny OD3}} & 0\\
\mathscr{J}_{\text{\tiny D4D3}}^\ast&0&\mathscr{J}_{\text{\tiny D5D4}}&\mathscr{J}_{\text{\tiny D6D4}}&\mathscr{J}_{\text{\tiny OD4}} & 0\\
\mathscr{J}_{\text{\tiny D5D3}}^\ast&
\mathscr{J}_{\text{\tiny D5D4}}^\ast&0&\mathscr{J}_{\text{\tiny D6D5}}&\mathscr{J}_{\text{\tiny OD5}}& 0\\
\mathscr{J}_{\text{\tiny D6D3}}^\ast&
\mathscr{J}_{\text{\tiny D6D4}}^\ast&\mathscr{J}_{\text{\tiny D6D5}}^\ast&0&\mathscr{J}_{\text{\tiny OD6}}& 0\\
\mathscr{J}_{\text{\tiny OD3}}^\ast&
 \mathscr{J}_{\text{\tiny OD4}}^\ast & \mathscr{J}_{\text{\tiny OD5}}^\ast& \mathscr{J}_{\text{\tiny OD6}}^\ast&0 &\hbar \frac{\Omega_\mu^\ast}{2} \\
 0&0&0&0 &\hbar \frac{\Omega_\mu}{2} & 0
\end{array}
\right),\label{sec06-eq3}
\end{eqnarray}
in the basis of $ \{ \lvert \uparrow_{\text{a}},e,\uparrow,\uparrow,\uparrow\rangle,\lvert \uparrow_{\text{a}},\uparrow,e,\uparrow,\uparrow\rangle,\lvert \uparrow_{\text{a}},\uparrow,\uparrow,e,\uparrow\rangle ,   \lvert \uparrow_{\text{a}},\uparrow,\uparrow,\uparrow,e\rangle ,     \lvert e_{\text{a}},\uparrow,\uparrow,\uparrow,\uparrow\rangle     ,  \lvert \downarrow_{\text{a}},\uparrow,\uparrow,\uparrow,\uparrow\rangle  \} $.

\subsection{Residual DDI between data molecules}\label{sec06b}
A perfect ancilla-assisted gate is based on zero DDI between any two data molecules. For the layout in Fig.~\ref{figure-6m}, there is indeed no DDI between the molecules at $D_4$ and $D_6$, or between those at $D_3$ and $D_5$. But the interactions between these two groups are nonzero. To analyze the residual DDI, we note that the two sites $D_4$ and $D_5$ are symmetric to each other with respect to $O$, and similarly for the two sites $D_3$ and $D_6$. We can analyze the DDI between $D_4$ and $D_3$ and that between $D_4$ and $D_5$ as an example.

For the layout of the molecules in Fig.~\ref{figure-6m}, we use $L_2$ to denote the distance between $O$ and $P_2$. One can find that the DDI between the ancilla at $O$ and the data molecule at $D_4$ is given by
\begin{eqnarray}
\mathscr{J}_{\text{\tiny OD4}}&=&-\frac{1}{4\pi\epsilon_0}\frac{\text{\textdong}^2}{2|OD_4|^3}\sin^2\angle D_4OP_2\nonumber\\
&=& - \frac{s^2}{(1-2s+3s^2)^{5/2}}\mathbb{J}^\prime,\label{4qubit-01}
\end{eqnarray}
where $s \equiv |D_4P_2|/|P_1P_2|\in(0,1)$ and
\begin{eqnarray}
\mathbb{J}^\prime=\frac{1}{4\pi\epsilon_0}\frac{\text{\textdong}^2}{L_2^3}.
\end{eqnarray}
The residual DDI between the data molecules at $D_4$ and $D_5$ is given by
\begin{eqnarray}
\mathscr{J}_{\text{\tiny D4D5}}&=&-\frac{1}{4\pi\epsilon_0}\frac{\text{\textdong}^2}{6|D_4D_5|^3}(1-3\cos^2\angle D_4OP_2)\nonumber\\
&=& \frac{1-2s}{24(1-2s+3s^2)^{5/2}}\mathbb{J}^\prime,\label{4qubit-02}
\end{eqnarray}
which shows that the interaction is positive when $s<0.5$, zero when $s=0.5$, and negative when $0.5<s<1$. The magnitude of $\mathscr{J}_{\text{\tiny OD4}}$ and the value of $\mathscr{J}_{\text{\tiny D4D5}}$ are shown by the solid and dashed curves in Fig.~\ref{figure-6m}(b) in units of $\mathbb{J}^\prime$. There is also DDI between the molecules at $D_3$ and $D_4$, or between the molecules at $D_5$ and $D_6$, which can be studied with a similar procedure as above.

\subsection{Fidelity of the three and four-qubit controlled phase gates}\label{sec06c}
To demonstrate that the ancilla-assisted mechanism is useful to realize multi-qubit gate, we consider the three and four-qubit gates in more detail. For the four-qubit gate, the four ancilla in Fig.~\ref{figure-6m}(a) are considered. For the three qubit gate, the ancilla at sites $D_3,D_5$, and $D_6$ are used for the analyses. This is because the undesired DDI between molecules at $D_4$ and $D_5$ is stronger compared to other unwanted DDI in the setup, so that a gate that avoids it can have a higher fidelity. This means that sites at $(D_3,D_4,D_6)$ can have the same effect as $(D_3,D_5,D_6)$.

In order to have stronger DDI between the data molecules and the ancilla, it is desirable to trap the molecules close enough. For this purpose, it is useful to consider the layout shown in Fig.~\ref{figure-6m}(a). To simplify the analysis, we can choose a condition when the interaction $\mathscr{J}_{\text{\tiny OD}_j}$ are the same for $j\in\{3,4,5,6\}$. We find that when
\begin{eqnarray}
|D_4P_2|\approx 0.3368|P_1P_2|,~|D_4D_6|= \sqrt{6}|OP_2|/3,\label{condition}
\end{eqnarray}
the interactions $\mathscr{J}_{\text{\tiny OD4}}=\mathscr{J}_{\text{\tiny OD6}}\approx-0.3125\mathbb{J}^\prime$, and meanwhile the interactions between the pairs of the data molecules symmetric to each other with respect to $O$ are $\mathscr{J}_{\text{\tiny D4D5}} \approx0.03747\mathbb{J}^\prime$ and $\mathscr{J}_{\text{\tiny D3D6}} \approx-0.01228\mathbb{J}^\prime$, and $\mathscr{J}_{\text{\tiny D3D4}} \approx0.005677\mathbb{J}^\prime$. The choice of the distance between $D_4$ and $D_6$ to be $\sqrt{6}|OP_2|/3$ is to approximately ensure that the distance between $D_4$ and $O$, equal to about $0.8165 |OP_2|$, is very near to $|D_4D_6|$. At this condition, the distance between $O$ and $D_6$ is about $1.159|OP_2|$. 

\begin{figure}
\includegraphics[width=5.5in]
{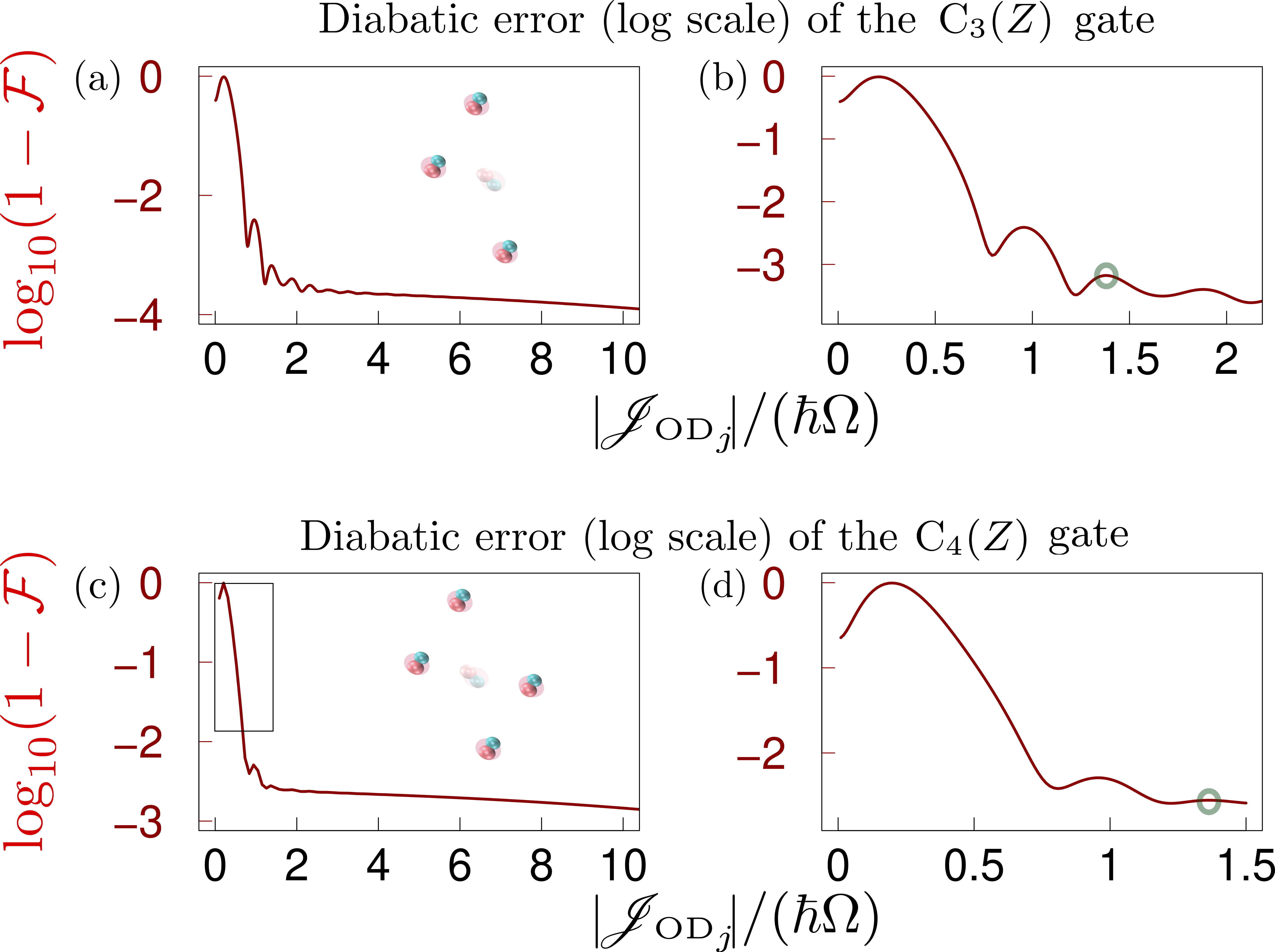}
\caption{(a,b) and (c,d) show the log$_{10}(1-\mathcal{F})$ of the three and four-qubit controlled phase gates, respectively. The simulation is with the pulse in Eq.~(\ref{omega-time}), which corresponds to a $\text{C}_3(Z)$ gate in (a,b) and a $\text{C}_4(Z)$ gate in (c,d). $\mathscr{J}_{\text{\tiny OD}_j}$ is the DDI between the ancilla and one data molecule, which is equal for all the data molecules with the condition detailed around Eq.~(\ref{condition}). The DDI between the data molecules are proportional to $\mathscr{J}_{\text{\tiny OD}_j}$, with coefficients given below Eq.~(\ref{condition}). The circle in (b) indicates a point when the fidelity of the $\text{C}_3(Z)$ gate has a local minimum of $0.9993$ at $|\mathscr{J}_{\text{\tiny OD}_j}|/(\hbar\Omega)\approx1.38$. Similarly, the circle in (d) indicates a similar point when $\mathcal{F}= 0.9972$ at $|\mathscr{J}_{\text{\tiny OD}_j}|/(\hbar\Omega)\approx1.36$. These two points are near to the condition used in Fig.~\ref{figure-dyn-error4} for showing the state dynamics for each class of input eigenstates.
\label{figure-uni-error4} }
\end{figure}

\begin{figure}
\includegraphics[width=5.5in]
{dynamics-4qubit.png}
\caption{State dynamics in the $\text{C}_4(Z)$ gate with the molecular configuration in Fig.~\ref{figure-6m} and $|\mathscr{J}_{\text{\tiny OD}_j}|/(\hbar\Omega)\approx1.37$. The gate time is $T\approx2.94$~ms with DDI listed in Eq.~(\ref{DDI-example}). $\delta P$ shows the finial population error in the computational basis state for each panel. The solid and dashed curves in (a), (b1-b4), (c1-c2), and (d) show the population and phase evolution in the computational basis states when the input state of the gate is the computational basis state of class (ii), (iii), (iv), and (v), respectively. The y-axis on the left~(right) of each panel is for population~(phase). The molecular configuration shown to the left of panels (a,b2,b4,c2), or to the right of panels (b1,b3,c1,d) indicate the input states used in the simulation for the corresponding panel, where each of the shown sites hosting a data molecule has the molecule initialised in $\uparrow$. Because $|\mathscr{J}_{\text{\tiny OD}_j}|$ is equal for $j\in\{3,4,5,6\}$, the four types of input states in class (ii) have similar dynamics. So, we show in (a) a representative result with an input state with the molecule at site $D_3$ initialised in $\uparrow$. Each of the four states in class (iii) involves a specific DDI between the two data molecules that are initially in the state $\uparrow$, so we have four cases shown in (b1-b4). Similarly, there are four input states in class (iv), but the DDI involved only has two species, shown by the representative molecular configuration accompanying panels (c1) and (c2). Panel (d) shows results when all the four data molecules are initialised in $\uparrow$. 
\label{figure-dyn-error4} }
\end{figure}

To show that the configuration of the molecular locations in Fig.~\ref{figure-6m} can yield kHz-scale DDI between the ancilla and each of the data molecules, and meanwhile possessing negligible DDI between any data molecules, we consider NaCs as in Ref.~\cite{picard_entanglement_2025}, and list one example of the numerical data for the interactions between the molecules at the sites when $\{|OP_2,|OD_4|,|OD_6| \}\approx\{1.26,1.03,1.46\}~\mu$m,
\begin{eqnarray}
O \text{\textemdash}D_{j}:&&~~~~h\times(-1)~\text{kHz},~~j\in\{3,4,5,6\},\nonumber\\
D_3 \text{\textemdash}D_4:&&~~~~h\times0.0182~\text{kHz},~~(\text{same for the molecules at }D_5 \text{\textemdash}D_6),\nonumber\\
D_3 \text{\textemdash}D_6:&&~~~~h\times(-0.0393)~\text{kHz},\nonumber\\
D_4 \text{\textemdash}D_5:&&~~~~h\times0.120~\text{kHz},\label{DDI-example}
\end{eqnarray}
and the interaction between molecules at $D_3$ and $D_5$ is zero, similarly for the two sites $D_4$ and $D_6$, because of the condition $\angle OP_2P_1=$arccos$(1/\sqrt{3})$ and $D_3D_5//D_4D_6$. Here, because the two lines $D_3D_4$ and $D_5D_6$ are parallel to each other, and $|D_3D_4|=|D_5D_6|$, the DDI between the two data molecules at $D_3$ and $D_4$ is equal to that between the two data molecules at $D_5$ and $D_6$. Note that the result shown in Figs.~\ref{figure-uni-error4} and~\ref{figure-dyn-error4} does not depend on the numerical value of these DDI, but only on the ratio between DDI and $\Omega$. 

With the ratio between the DDI at hand, we can numerically study the diabatic error of the three and four-qubit controlled phase gate. Because a controlled phase $\pi$ is of particular interest for quantum error correction~\cite{Shi2018prapp}, we consider the case of $\pi$, i.e., the $\text{C}_3(Z)$ and $\text{C}_4(Z)$ gates. For a numerical simulation of the gate fidelity, the states in class (ii), (iii), (iv) and (v) are studied with the Hamiltonians in Eq.~(\ref{adia-01}), Eq.~(\ref{duu}), Eq.~(\ref{sec06-eq2}), and Eq.~(\ref{sec06-eq3}), respectively. Note that when these Hamiltonians are used, the actual DDI should be correctly set according to each state in each class of the states. The gate error as a function of the ratio between $\mathscr{J}_{\text{\tiny OD4}}$ and $\hbar\Omega$ is shown in Fig.~\ref{figure-uni-error4}. Due to the presence of the DDI between data molecules that partially breaks the dark-state condition, the error is larger than that in the $\text{C}_2(Z)$ gate shown in Fig~\ref{figure-uni-error}. Different from Fig~\ref{figure-uni-error}, the fidelity in $\text{C}_3(Z)$ and $\text{C}_4(Z)$ nearly increases monotonically when $|\mathscr{J}_{\text{\tiny OD4}}|$ increases beyond $2\hbar\Omega$. The data in Fig.~\ref{figure-uni-error4} shows that $\mathcal{F}>0.999~(0.99)$ when $|\mathscr{J}_{\text{\tiny OD4}}|/(\hbar\Omega)>1.14~(0.708)$ for the C$_3(Z)$~[C$_4(Z)$] gate.

Because of the nonzero DDI between the data molecules, the dark-state condition can fail. To study how the diabatic effect as well as the dynamical phase affect the performance of the gate, we show the population and phase dynamics in Fig.~\ref{figure-dyn-error4} for each representative states in the $\text{C}_4(Z)$ gate. Because the state dynamics involves time evolution, we choose, for example, $|\mathscr{J}_{\text{\tiny OD}_j}|/(\hbar\Omega)\approx1.37$ when the ratio between DDIs are according to those in Eq.~(\ref{DDI-example}). The reason to choose $|\mathscr{J}_{\text{\tiny OD}_j}|/(\hbar\Omega)\approx1.37$ is because the circle in Fig.~\ref{figure-uni-error4}(b) indicates a point when the fidelity of the $\text{C}_3(Z)$ gate has a local minimum at $|\mathscr{J}_{\text{\tiny OD}_j}|/(\hbar\Omega)\approx1.38$, and the $\text{C}_4(Z)$ gate has a locally minimal fidelity at $|\mathscr{J}_{\text{\tiny OD}_j}|/(\hbar\Omega)\approx1.36$ as indicated by the circle in Fig.~\ref{figure-uni-error4}(d). Then, $|\mathscr{J}_{\text{\tiny OD}_j}|/(\hbar\Omega)\approx1.37$ can nearly put both $\text{C}_3(Z)$ and $\text{C}_4(Z)$ into a regime of being robust against small fluctuation of DDI, so that Fig.~\ref{figure-dyn-error4} is shown for the $\text{C}_4(Z)$ gate, but also applies to the $\text{C}_3(Z)$ gate. The final population leakage outside of the computational basis state is highlighted as $\delta P$ for each panel of Fig.~\ref{figure-dyn-error4}. One can see that the state dynamics in panel (b2) matches the dark-state picture best. This is because for the input state of panel (b2), Eq.~(\ref{duu}) reduces to Eq.~(\ref{adia2-01}) when the fluctuation of DDI is ignored, as considered here. Then, the state dynamics for panel (a) and panel (b2) are governed by the same form of Hamiltonian, with the DDI in panel (b2) equal to $\sqrt{2}$ times the DDI in panel (a). The $\sqrt{2}$-fold stronger DDI can yield a stronger dark-state condition, which is why the population error in Fig.~\ref{figure-dyn-error4}(b2) is smaller than Fig.~\ref{figure-dyn-error4}(a), whilst both panels have exactly zero phase in the computational basis during the pulse. Here the Hamiltonian is real and the Rabi frequency retraces, yielding a zero geometric phase. In the other panels, because of the breakdown of the dark-state condition, nonzero dynamic phases arise. However, one can see that the major error is from the population loss. This is why the $\text{C}_3(Z)$ gate has a larger fidelity compared to the $\text{C}_4(Z)$ gate because in the results of Fig.~\ref{figure-uni-error4}(a,b) for the $\text{C}_3(Z)$ gate, the initial molecular state corresponds to the one used in Fig.~\ref{figure-dyn-error4}(c1). The population error in Fig.~\ref{figure-dyn-error4}(c1) is much smaller than that in Fig.~\ref{figure-dyn-error4}(c2), alongside the absence of the state evolution in panel (d) which also has a large population error, leading to a larger gate fidelity in the $\text{C}_3(Z)$ gate of Fig.~\ref{figure-uni-error4}.

Due to the much larger diabatic error in the $\text{C}_3(Z)$ and $\text{C}_4(Z)$ gates as compared to the $\text{C}_2(Z)$ gate, and the error from the polarization impurity in the microwave field and QM is of similar magnitude to that of diabatic error in the $\text{C}_2(Z)$ gate, the error from these latter effect should be much smaller than the diabatic error of the $\text{C}_3(Z)$ and $\text{C}_4(Z)$ gates. Therefore, we do not study QM and polarization impurity for the $\text{C}_3(Z)$ and $\text{C}_4(Z)$ gates.


\section{Discussions}\label{sec07}
The dark-state gate in this work depends on an ancilla-data system, where one ancilla interacts with all the data qubits, but no interaction should occur between any two data molecules. The gate is executed by microwave field which covers all the molecules, therefore it is necessary to choose different levels for the ancilla as compared to the data molecules. For this purpose, one example is shown in Fig.~\ref{figure-hybrid} with $^{23}$Na$^{133}$Cs. Other molecules are also useful once a resonant DDI can occur between the ancilla and the data molecule.

For the multi-qubit gates with more than two data molecules, it is necessary to configure the data molecules spatially so that no DDI occur among them. We show an example with a minimal four data qubits in Fig.~\ref{figure-6m} which can approximately fulfill this condition, but it is possible to have more data qubits by numerically searching optimal spatial arrangement for the molecules as well as using pulse shaping technique~\cite{Lu_2026,bergonzoni_iswap_2025} to suppress the diabatic leakage.

An alternative way to realize ancilla-assisted gates with many qubits is by moving molecules close to the ancilla before the gate, and then move away after the gate so as to allow further execution in the quantum circuit. Moving optically trapped molecules is possible, as in the experiments of Ref.~\cite{picard_entanglement_2025} via optical tweezer shuttling. The question is whether we can move many molecules simultaneously so as to avoid long gap time in the quantum circuit. In this case, all the six data molecules at sites $D_j$ with $j=1,\cdots,6$ in Fig.~\ref{figure-1d} can be used as data molecules, so that at least a six-qubit controlled-phase gate is realizable. The feasibility to realize such multi-qubit controlled gates relies on the absence of single-site addressing during the gate while preserving robustness against the quantized motion of the optically trapped molecules.

\section{Conclusions}\label{sec08}
We present an efficient multi-qubit controlled-phase gate using optically trapped ultracold polar molecules. The gate is realized via a dark-state mechanism driven by one or two microwave pulses. A dark state forms when the microwave field is resonant with an ancilla molecule, while a resonant dipole--dipole interaction (DDI) couples the ancilla to the data molecules. We illustrate a suitable level scheme using $^{23}\text{Na}^{133}\text{Cs}$ molecules possessing a vanishing rotational Zeeman effect; however, this protocol is broadly applicable to molecules where the rotational Zeeman effect is small compared to the nuclear spin Zeeman effect. The gate is remarkably resilient against both the quantized motion of molecules in optical traps and reasonable polarization impurities in the microwave fields. For a two-qubit controlled-phase gate, a $4\text{~ms}$ pulse yields a fidelity of approximately $0.996$, where most of the error is from qubit $T_2^\ast$ that is assumed to be $1.5$~s. Although the protocol is subject to diabatic errors, numerical simulations demonstrate that the diabatic error remains on the order of $10^{-4}$ for two-qubit and three-qubit gates, and $10^{-3}$ for a four-qubit gate, all assuming a modest DDI less than $25\hbar/T$, which is shallow in the adiabatic regime. These results establish the ancilla-assisted dark-state mechanism as a versatile control tool in ultracold polar molecules.

\section*{acknowledgments}We acknowledge the National Natural Science Foundation of China under Grants No. 12074300 and No. 12547103, thank the anonymous referees for valuable suggestions, and the Beijing Super Cloud Center for providing HPC resources.

\begin{appendix}

\begin{figure}
\includegraphics[width=3.30in]
{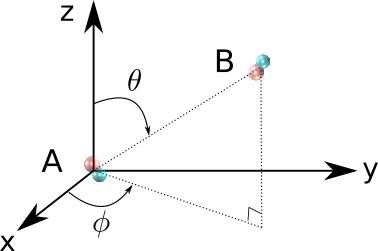}
\caption{Two polar molecules labeled A and B with relative orientation shown by the spherical angles $\theta$ and $\phi$, where z is the quantization axis. \label{figure-two-molecules} }
\end{figure}
\section{DDI of two polar molecules}\label{app-ddi}
The interaction between the dipole moments of two molecules can be calculated in a way similar to that for Rydberg atoms. DDI does not change the nuclear spin, so we omit the nuclear spin states here. Following Sec.~2.3.1 of Ref.~\cite{Shi2021qst}, we consider molecular states labeled with rotational quantum number $N_\alpha$ and its projection $m_{\alpha}$ onto the quantization axis, where $\alpha$ is the index for two molecules, A and B. We assume that the quantization axis is along $\mathbf{z}$. The dipole-dipole interaction between the two molecules with quantum numbers $(N_{\text{A}}, m_{\text{\tiny A}})$ and $(N_{\text{B}}, m_{\text{\tiny B}})$ is an electrostatic interaction given by
\begin{eqnarray}
\hat {V}_{\text{dd}} &=& \mathbf{\hat s}_{\text{A}} \cdot \left( 3\mathbf{\hat r}\mathbf{\hat r}/r^2-\mathbf{\hat I} \right) \cdot  \mathbf{\hat s}_{\text{B}}/(4\pi\epsilon_0 R^3),
\end{eqnarray}
where $\mathbf{\hat s}_{{\text{A(B)}}}$ is the dipole moment of molecule A(B), $R$ is the distance between the two molecules, $\mathbf{\hat I}$ is the identity operator, and $\epsilon_0$ is the dielectric permittivity in free space. Using the representation of spherical harmonic rank 2 tensor, the matrix element between the initial state and a DDI-coupled state  with quantum numbers $(N_{\text{a}}, m_{\text{\tiny a}})$ and $(N_{\text{b}}, m_{\text{\tiny b}})$
\begin{eqnarray}
&&V_{ m_{\text{A}} m_{\text{B}};m_{\text{a}} m_{\text{b}} }^{ N_{\text{A}} N_{\text{B}};N_{\text{a}} N_{\text{b}}} =\langle N_{\text{A}} m_{\text{\tiny A}}; N_{\text{B}} m_{\text{\tiny B}}  |\hat {V}_{\text{dd}} |   N_{\text{a}} m_{\text{\tiny a}}; N_{\text{b}} m_{\text{\tiny b}}  \rangle \label{eq101401}
\end{eqnarray}
can be written as
\begin{eqnarray}
&&-\frac{\sqrt6}{4\pi\epsilon_0R^3}\frac{\langle N_{\text{A}} ||D_{\text{A}}^{[1]}||     N_{\text{a}}\rangle \langle N_{\text{B}} ||D_{\text{B}}^{[1]}||     N_{\text{b}} \rangle }{\sqrt{(2N_{\text{A}}+1 )(2N_{\text{B}}+1 )}} \sum_{M=-2}^2\Psi_{-M}^{[2]}(\theta,\phi) \sum_{\alpha,\beta=-1}^1 \mathcal{C}_{\alpha \beta M}^{112} \mathcal{C}_{m_{\text{a}}\alpha  m_{\text{A}}}^{N_{\text{a}} 1 N_{\text{A}}} \mathcal{C}_{m_{\text{b}}\beta  m_{\text{B}}}^{N_{\text{b}} 1 N_{\text{B}}}  ,  \label{eq101402}
\end{eqnarray}
where $D_k^{[1]}$ is the dipole moment of atom $k=A$ or $B$ in terms of rank-1 tensor, $\mathcal{C}$ is a Clebsh-Gordan coefficient, $\Psi^{[2]}$ is a rank-2 tensor given by the standard spherical harmonics multiplied times $\sqrt{4\pi/5}$, and $(\theta,\phi)$ gives the angular position of molecule B with respect to molecule A as schematically shown in Fig.~\ref{figure-two-molecules}.

\subsection{Exchange DDI}\label{ex-DDI}
A usual case studied in literature is the exchange process that the $N$ and its projection onto the quantization axis exchange between the two molecules. Consider a case with $(N_{\text{A}}, m_{\text{\tiny A}})$ and $(N_{\text{B}}, m_{\text{\tiny B}})$ changed to $(N_{\text{b}}, m_{\text{\tiny b}})$ and $(N_{\text{a}}, m_{\text{\tiny a}})$, respectively, with change of angular momentum $\pm1$ for each molecule and $\Delta N=\pm1$. In the case of Ref.~\cite{picard_entanglement_2025}, the exchange between $(N_{\text{A}}, m_{\text{\tiny A}})=(0,0)$ and a superposition of $(N_{\text{B}}, m_{\text{\tiny B}})=(1,\pm1)$ was used, where $\langle N_{\text{A}} ||D_{\text{A}}^{[1]}||     N_{\text{a}}\rangle \langle N_{\text{B}} ||D_{\text{B}}^{[1]}||     N_{\text{b}} \rangle$=-\textdong$^2$. If $(N_{\text{B}}, m_{\text{\tiny B}})=(1,\pm1)$, the last sum in Eq.~(\ref{eq101402}) is equal to $\sqrt{2}/6$. With the factor $\frac{1}{\sqrt{(2N_{\text{A}}+1 )(2N_{\text{B}}+1 )}}$, this yields $\frac{1}{4\pi\epsilon_0R^3}\frac{\text{\textdong}^2}{6}(3\cos^2\theta-1)$. Because the actual state that was used in Ref.~\cite{picard_entanglement_2025} is an equal (anti-phased) superposition of $(N_{\text{B}}, m_{\text{\tiny B}})=(1,\pm1)$, the actual DDI coefficient should be enhanced by a factor of $\sqrt{2}$, so that the interaction is given by $-\frac{1}{4\pi\epsilon_0R^3}\frac{\sqrt{2}\text{\textdong}^2}{6}(1-3\cos^2\theta)$.

If we consider the state pairs $(N_{\text{A}}, m_{\text{\tiny A}})=(0,0)$ and $(N_{\text{B}}, m_{\text{\tiny B}})=(1,0)$, then the only difference is that the last sum in Eq.~(\ref{eq101402}) is equal to $-\sqrt{2}/3$, while the product of the dipole moments is -\textdong$^2$, so that the interaction coefficient is $\frac{1}{4\pi\epsilon_0R^3}\frac{ \text{\textdong}^2}{3}(1-3\cos^2\theta)$, which is the DDI appeared in most literature and in usual experimental setups~\cite{Cornish_2024,yu2026}.

\subsection{DDI in this article}
The interaction between two data molecules can be nonzero, in the form of spin exchange $\lvert\uparrow ,e\rangle \leftrightarrow\lvert e,\uparrow\rangle $. As shown above, the strength is $-\frac{1}{4\pi\epsilon_0R^3}\frac{\text{\textdong}^2}{6}(1-3\cos^2\theta)$.

The interaction between the ancilla and one data molecule is resonant with the following process,
\begin{eqnarray}
\lvert\uparrow_{\text{a}},e\rangle\leftrightarrow \lvert e_{\text{a}},\uparrow\rangle
\end{eqnarray}
where the states are listed in Eq.~(\ref{phy-01}). The process $\lvert\uparrow_{\text{a}},e\rangle\rightarrow \lvert e_{\text{a}},\uparrow\rangle$ decreases the total $m$ by two, and one can find
\begin{eqnarray}
V_{ 01;-10}^{ 01;10}
&=& \frac{\text{\textdong}^2}{8\pi\epsilon_0R^3}  \sin^2\theta e^{-2i\phi} . \label{ancillaDDI}
\end{eqnarray}

\section{Leakage transitions in the dark-state gate}\label{app-leakage}
The four states of Eq.~(\ref{phy-01}) will enable DDI. For the quantum gate, the other two states that are involved in the gate in this work can be
\begin{eqnarray}
\lvert\downarrow_{\text{a}}\rangle &= & \lvert2,-2,-3/2,-7/2\rangle,\nonumber\\
\lvert\downarrow\rangle &(\text{mainly:)  }& \lvert0,0,1/2,7/2\rangle,\label{app-phy-02a}
\end{eqnarray}
where $\lvert \downarrow\rangle$ and $\lvert \uparrow\rangle$ are chosen as the states $\lvert 1\rangle$ and $\lvert 0\rangle$ of Ref.~\cite{Ni_2018}. Among the two states above,  $\lvert \downarrow\rangle$ has multiple components due to hyperfine interaction, but this state is far detuned since the microwave field is for the transition between the $N=1$ and the $N=2$ states. Because all the spin projections are maximal in $\lvert2,-2,-3/2,-7/2\rangle$, hyperfine interaction does not mix it with other states~\cite{PhysRevA.107.023102}, the state $\lvert\downarrow_{\text{a}}\rangle$ is purely $\lvert2,-2,-3/2,-7/2\rangle$. If only the energy gap is concerned, then the microwave field is resonant for both $\lvert e_{\text{a}}\rangle\leftrightarrow\lvert\downarrow_{\text{a}}\rangle$ and
\begin{eqnarray}
\lvert e\rangle &\leftrightarrow & \lvert2, 2,3/2,7/2\rangle\equiv\lvert\text{leak},e+\rangle, \label{app-e-leak01}
\end{eqnarray}
where the state on the right side is denoted as $\lvert\text{leak},e+\rangle$ because the transition in Eq.~(\ref{app-e-leak01}) needs a $\sigma^+$ polarized field. When the desired transition $\lvert e_{\text{a}}\rangle\leftrightarrow\lvert\downarrow_{\text{a}}\rangle$ is driven by a $\sigma^-$ polarized field, the leakage in Eq.~(\ref{app-e-leak01}) is suppressed unless the field has some polarization impurity.

\subsection{Leakage when the polarization of the field is perfect}
However, for the $\sigma^-$ polarized field, the state $\lvert e\rangle$ can go to a state containing the component $\lvert2, 0,3/2,7/2\rangle$. This state component is not maximally stretched, therefore hyperfine interaction can couple it with other states. We find that at $B=2$~G, there are two states with a population larger than 0.1 in this state component. Among these two states, one can be neglected if $\Omega$ of order of 1~kHz is used because that state is detuned from the microwave transition by about $14.4$~kHz, while the other state is detuned only by about 3~kHz. When keeping state components with population larger than 0.001, this latter state is
\begin{eqnarray}
&&\lvert\text{leak},e-\rangle=\beta_0^{(e-)}\lvert2,0,3/2,7/2\rangle+ \beta_1^{(e-)}\lvert2,1,3/2,5/2\rangle+ \beta_2^{(e-)}\lvert2,1,1/2,7/2\rangle\nonumber\\
&&~~~~~~~~~~~ +  \beta_3^{(e-)}\lvert2,2,-1/2,7/2\rangle +  \beta_4^{(e-)}\lvert2,2,3/2,3/2\rangle +   \beta_5^{(e-)}\lvert2,2,1/2,5/2\rangle +\lvert o\rangle, \label{app-phy-02}\\
&&\{|\beta_0^{(e-)}|^2 ,~|\beta_1^{(e-)}|^2,~|\beta_2^{(e-)}|^2,~|\beta_3^{(e-)}|^2,~|\beta_4^{(e-)}|^2,~|\beta_5^{(e-)}|^2\}\approx\{0.5968,~0.2448,~0.0790,~0.0428,\nonumber\\
&&~~~~~0.0337,~0.0029\},\nonumber
\end{eqnarray}
where e- in $\lvert\text{leak},e-\rangle$ means that the leakage is from $\lvert e\rangle$ and is via the $\sigma^-$ polarized microwave field, and $\lvert o\rangle$ is a state with negligible population. In the equation above, the phase is zero in the coefficients $\beta_j^{(e-)}$ for all $j\in\{0,\cdots,5\}$. The transition $\lvert e\rangle \leftrightarrow \lvert\text{leak},e-\rangle$ is detuned from $\lvert e_{\text{a}}\rangle \leftrightarrow \lvert\downarrow_{\text{a}} \rangle$ by 
\begin{eqnarray}
\Delta^{(e-)}/(2\pi)&=&-3.13~\text{kHz}.\label{e-Delta}
\end{eqnarray}
Here, the detuning is given by the frequency of the microwave field minus the transition frequency. Due to the angular momentum coupling, there is a factor of $1/\sqrt{6}$ with the leakage transition, so that if the Rabi frequency for the desired transition $\lvert e_{\text{a}}\rangle \leftrightarrow \lvert\downarrow_{\text{a}} \rangle$ is $\Omega_\mu$, then the Rabi frequency for the transition $\lvert e\rangle \leftrightarrow \lvert\text{leak},e-\rangle$ would be 
\begin{eqnarray}
\Omega_{e-} &=&   \beta_0^{(e-)}\Omega_\mu/\sqrt{6} .\label{e-omega}
\end{eqnarray}

The reason that we choose B=2~G in the analysis is as follows. Because the theory in this work can be used with a molecule with a negligible $g_r$ in a Gauss-scale magnetic field, we take a B-field of 2~G as an example in this work. Note that for the case of $^{23}$Na$^{133}$Cs, a strong B-field is applicable since $g_r=0$.

Before giving the details about other possible leakage channels, we first preclude some channels that are largely detuned, and, hence, can be neglected. The microwave field is for a transition between $N=1$ and $N=2$, so that the state $\lvert\downarrow\rangle$ with $N=0$ is far detuned, thus does not lead to leakage. Similarly for the states $\lvert\uparrow_{\text{a}}\rangle$ and $\lvert\uparrow\rangle$ because they are in the $N=0$ manifold.

The above analysis is based on perfect polarization in the microwave field. When there is a  wrong component in the polarization, other leakage channels can be opened.

\subsection{Leakage with impure polarization in the microwave field}
If the input of the gate is coupled with DDI, then the state $\lvert e_{\text{a}}\rangle$ is barely populated according to the dark-state mechanism, but DDI will cause the state $\lvert e\rangle$ to be populated. On the other hand, if the input is not coupled with DDI, then $\lvert e_{\text{a}}\rangle$ can be populated. We study these two cases below.

\subsubsection{Impure polarization on the state $\lvert e\rangle$}
When DDI occurs, the state $\lvert e\rangle$ can be populated from the exchange process, and from it there can be some leakage if there is polarization impurity in the field. We can write the polarization vector of the microwave field as
\begin{eqnarray}
\mathbf{e}_{\mu} &=& \sum_{\alpha\in\{0,\pm\}} \iota_\alpha \mathbf{e}_{\alpha},
\end{eqnarray}
where $ \sum_{\alpha\in\{0,\pm\}} |\iota_\alpha|^2=1$, $\mathbf{e}_{\pm}= (\mathbf{x}\mp i\mathbf{y})/\sqrt{2}$, and $\mathbf{e}_0=\mathbf{z}$. For the $\sigma^+$ polarized field, $\lvert e\rangle$ is driven to $\lvert2,2,3/2,7/2\rangle$ as in Eq.~(\ref{app-e-leak01}). If the desired Rabi frequency is $\Omega_\mu$ for the correct transition with ideal polarization, then the wrong Rabi frequency for the excitation of $\lvert e\rangle$ as in Eq.~(\ref{app-e-leak01}) is
\begin{eqnarray}
\Omega_{e+} &=&  \iota_{+} \Omega_\mu ,\label{e+omega}
\end{eqnarray}
where there is no extra angular momentum coefficient because the wrong transition in Eq.~(\ref{app-e-leak01}) and the desired transition are symmetric to each other.

 For the $\pi$ polarized field, $\lvert e\rangle$ can be driven to a state with component $\lvert2,1,3/2,7/2\rangle$. We find that if restricting to states with population larger than 0.01, there are three such states, among which two are detuned by about 33.5~kHz and 13.2~kHz, and another one is $\lvert\text{leak},e0\rangle$, which is detuned by
\begin{eqnarray}
\Delta^{(e0)}/(2\pi)&=&- 1.06~\text{kHz}.\label{e0Delta}
\end{eqnarray}
Because the leakage is a higher order process, we only consider the last of these three states since it is not far detuned.
If keeping state components with population larger than 0.001 only, we have
\begin{eqnarray}
&&\lvert\text{leak},e0\rangle=\beta_0^{(e0)}\lvert2,2,3/2,5/2\rangle+ \beta_1^{(e0)}\lvert2,1,3/2,7/2\rangle+ \beta_2^{(e0)}\lvert2,2,1/2,7/2\rangle  , \label{app-phy-03}\\
&&\{|\beta_0^{(e0)}|^2 ,~|\beta_1^{(e0)}|^2,~|\beta_2^{(e0)}|^2 \}\approx\{0.4610,0.3059,0.2331\},\nonumber
\end{eqnarray}
where the three state coefficients are positive. With an angular momentum factor $1/\sqrt{2}$, the Rabi frequency for the excitation of $\lvert\text{leak},e0\rangle$ is
\begin{eqnarray}
\Omega_{e0} &=&   \Omega_\mu \beta_1^{(e0)}\iota_{0}/\sqrt{2}  .\label{e0omega}
\end{eqnarray}

\subsubsection{Impure polarization on the state $\lvert e_{\text{a}}\rangle$}\label{app-ea-leak}
For the input state $\lvert\downarrow_{\text{a}}, \downarrow,\downarrow\rangle$, no DDI occurs, and when the microwave field excites it back to $\lvert e_{\text{a}}, \downarrow,\downarrow\rangle$, impure polarization in the field can excite it to other states in the $N=2$ manifold. For the $\pi$ polarized field, $\lvert e_{\text{a}}\rangle$ can be excited to a state containing $\lvert 2,-1,-3/2,-7/2\rangle$. At $B=2$~G, there are three states with population in this state larger than 0.001. Among these three states, one is over $\lvert\downarrow_{\text{a}}\rangle$ by about 12.2~kHz, another below $\lvert\downarrow_{\text{a}}\rangle$ by about 37.2~kHz. For a Rabi frequency of order $2\pi\times1$~kHz, these two states can be ignored. The third state is only detuned by
\begin{eqnarray}
\Delta^{(e_{\text{a}}0)}/(2\pi)&=&   1.04~\text{kHz}.\label{ea0Delta}
\end{eqnarray}
If keeping state components with population larger than 0.001 only, this state is given by
\begin{eqnarray}
&&\lvert\text{leak},e_{\text{a}}0\rangle=\beta_0^{(e_{\text{a}}0)}\lvert2,-2,-3/2,-5/2\rangle+ \beta_1^{(e_{\text{a}}0)}\lvert2,-1,-3/2,-7/2\rangle\nonumber\\
&&~~~~~~~~~~~~~~~~~+ \beta_2^{(e_{\text{a}}0)}\lvert2,-2,-1/2,-7/2\rangle , \label{app-phy-19}\\
&&\{|\beta_0^{(e_{\text{a}}0)}|^2 ,~|\beta_1^{(e_{\text{a}}0)}|^2,~|\beta_2^{(e_{\text{a}}0)}|^2 \}\approx\{0.5367,0.2663,0.1970\},\nonumber
\end{eqnarray}
where all the three coefficients are positive. With an angular momentum factor $1/\sqrt{2}$, the Rabi frequency for exciting $\lvert e_{\text{a}}\rangle$ to $\lvert\text{leak},e_{\text{a}}0\rangle$ is
\begin{eqnarray}
\Omega_{e_{\text{a}}0} &=&   \Omega_\mu \beta_1^{(e_{\text{a}}0)}\iota_{0}/\sqrt{2} . \label{ea0omega}
\end{eqnarray}

The $\sigma^+$ polarized field will excite the state $\lvert e_{\text{a}}\rangle$ to a state containing $\lvert 2,0,-3/2,-7/2\rangle$. At $B=2$~G, there are four states with population of $\lvert 2,0,-3/2,-7/2\rangle$ larger than 0.001, and among these four, only two have a population larger than 0.1. To get the leading error in the leakage, we can neglect the states with population smaller than 0.1. Within the two remaining states, one is over $\lvert\downarrow_{\text{a}}\rangle$ by about 11.2~kHz. For a Rabi frequency of order $2\pi\times1$~kHz, this state can be ignored, and the other is detuned by
\begin{eqnarray}
\Delta^{(e_{\text{a}}+)}/(2\pi)&=&  - 1.54~\text{kHz},\label{ea+Delta}
\end{eqnarray}
and is given by
\begin{eqnarray}
&&\lvert\text{leak},e_{\text{a}}+\rangle=\beta_0^{(e_{\text{a}}+)}\lvert2,0,-3/2,-7/2\rangle+ \beta_1^{(e_{\text{a}}+)}\lvert2,-1,-3/2,-5/2\rangle+ \beta_2^{(e_{\text{a}}+)}\lvert2,-2,-3/2,-3/2\rangle\nonumber\\
&&~~~~ +  \beta_3^{(e_{\text{a}}+)}\lvert2,-1,-1/2,-7/2\rangle +  \beta_4^{(e_{\text{a}}+)}\lvert2,-2,1/2,-7/2\rangle +   \beta_5^{(e_{\text{a}}+)}\lvert2,-2,-1/2,-5/2\rangle , \label{app-phy-20}\\
&&\{|\beta_0^{(e_{\text{a}}+)}|^2 ,~|\beta_1^{(e_{\text{a}}+)}|^2,~|\beta_2^{(e_{\text{a}}+)}|^2,~|\beta_3^{(e_{\text{a}}+)}|^2,~|\beta_4^{(e_{\text{a}}+)}|^2,~|\beta_5^{(e_{\text{a}}+)}|^2\}\approx\{0.4871,~0.2915,~0.1014,~0.0881,\nonumber\\
&&~~~~~0.0266,~0.0053\},\nonumber
\end{eqnarray}
when only keeping state components with population larger than 0.001. Its excitation has an angular momentum coupling factor $1/\sqrt{6}$, and the Rabi frequency is given by
\begin{eqnarray}
\Omega_{e_{\text{a}}+} &=&   \Omega_\mu \beta_0^{(e_{\text{a}}+)}\iota_{+}/\sqrt{6}.\label{ea+omega}
\end{eqnarray}

\subsection{Hamiltonian with leakage transitions}\label{app-leak-H}
With the leakage channels, the Hamiltonian for $ \lvert \downarrow_{\text{a}},\uparrow,\downarrow\rangle$ in Eq.~(\ref{adia-01}) should be updated to
\begin{eqnarray}
\hat{H}_{\text{ii}} &=&\hbar\left(
\begin{array}{cccccc}
  \Delta^{(e-)}& 0& 0&\frac{\Omega_{e-}}{2}& 0& 0\\
  0& \Delta^{(e0)}& 0&\frac{\Omega_{e0}}{2} & 0&  0\\
 0& 0&  0& \frac{\Omega_{e+}}{2}& 0&  0 \\
  \frac{\Omega_{e-}^\ast}{2}& \frac{\Omega_{e0}^\ast}{2} & \frac{\Omega_{e+}^\ast}{2} &0&\frac{\mathscr{J}}{\hbar}& 0\\
  0& 0& 0 &\frac{\mathscr{J}^\ast}{\hbar}&0 & \frac{\Omega_\mu^\ast}{2} \\
  0&0&  0&0 & \frac{\Omega_\mu}{2} & 0
\end{array}
\right)\label{adia-01-app}
\end{eqnarray}
where the former three states in the basis are $ \lvert \uparrow_{\text{a}}\rangle \otimes\lvert \text{leak},e-\rangle\otimes \lvert \downarrow\rangle ,~\lvert \uparrow_{\text{a}}\rangle \otimes\lvert \text{leak},e0\rangle\otimes \lvert \downarrow\rangle,~ \lvert \uparrow_{\text{a}}\rangle \otimes\lvert \text{leak},e+\rangle\otimes \lvert \downarrow\rangle$, and the latter three are $ \{    \lvert \uparrow_{\text{a}},e,\downarrow\rangle      ,  \lvert e_{\text{a}},\uparrow,\downarrow\rangle     ,  \lvert \downarrow_{\text{a}},\uparrow,\downarrow\rangle  \} $. Similarly, the Hamiltonian for $ \lvert \downarrow_{\text{a}},\uparrow,\uparrow\rangle$ in Eq.~(\ref{adia2-01}) should be updated to
\begin{eqnarray}
\hat{H}_{\text{iii}} &=&\hbar\left(
\begin{array}{cccccc}
  \Delta^{(e-)}& 0& 0&\frac{\Omega_{e-}}{2}& 0& 0\\
  0& \Delta^{(e0)}& 0&\frac{\Omega_{e0}}{2} & 0&  0\\
 0& 0&  0& \frac{\Omega_{e+}}{2}& 0&  0 \\
  \frac{\Omega_{e-}^\ast}{2}& \frac{\Omega_{e0}^\ast}{2} & \frac{\Omega_{e+}^\ast}{2} &0&\frac{A}{\hbar}& 0\\
  0& 0& 0 &\frac{A}{\hbar}&0 & \frac{\Omega_\mu^\ast}{2} \\
  0&0&  0&0 & \frac{\Omega_\mu}{2} & 0
\end{array}
\right)\label{adia-01-app}
\end{eqnarray}
where the latter three states in the basis are $ \{\lvert \Lambda \rangle      ,  \lvert e_{\text{a}},\uparrow,\uparrow\rangle     ,  \lvert \downarrow_{\text{a}},\uparrow,\uparrow\rangle  \} $, and $A$ and $\lvert \Lambda\rangle$ are defined around Eq.~(\ref{adia2-01}). The first three states in the basis of Eq.~(\ref{adia-01-app}) are
\begin{eqnarray}
 \text{The first: } && \frac{1}{A}\left( \mathscr{J}_{\text{ac}} \lvert \uparrow_{\text{a}}\rangle \otimes\lvert \text{leak},e-\rangle\otimes \lvert \uparrow\rangle   +\mathscr{J}_{\text{at}}  \lvert \uparrow_{\text{a}}\rangle \otimes \lvert \uparrow\rangle \otimes\lvert \text{leak},e-\rangle \right),\nonumber\\
 \text{The second: } && \frac{1}{A}\left( \mathscr{J}_{\text{ac}} \lvert \uparrow_{\text{a}}\rangle \otimes\lvert \text{leak},e0\rangle\otimes \lvert \uparrow\rangle   +\mathscr{J}_{\text{at}}  \lvert \uparrow_{\text{a}}\rangle \otimes \lvert \uparrow\rangle \otimes\lvert \text{leak},e0\rangle \right),\nonumber\\
 \text{The third: } && \frac{1}{A}\left( \mathscr{J}_{\text{ac}} \lvert \uparrow_{\text{a}}\rangle \otimes\lvert \text{leak},e+\rangle\otimes \lvert \uparrow\rangle   +\mathscr{J}_{\text{at}}  \lvert \uparrow_{\text{a}}\rangle \otimes \lvert \uparrow\rangle \otimes\lvert \text{leak},e+\rangle \right).
\end{eqnarray}

For the input state $\lvert \downarrow_{\text{a}},\downarrow,\downarrow\rangle$, there is no DDI, but only the microwave field will excite the ancilla state transitions
\begin{eqnarray}
 \lvert \downarrow_{\text{a}}\rangle  \leftrightarrow \lvert e_{\text{a}}\rangle,\nonumber\\
 \lvert \text{leak}, e_{\text{a}}0\rangle  \leftrightarrow \lvert e_{\text{a}}\rangle,\nonumber\\
 \lvert \text{leak}, e_{\text{a}}+\rangle  \leftrightarrow \lvert e_{\text{a}}\rangle,
\end{eqnarray}
where the latter two are from the impure polarization in the microwave fields. Because the states in the data molecules do not change, we can write the Hamiltonian for the ancilla only,
\begin{eqnarray}
\hat{H}_{\text{i}} &=&\hbar\left(
\begin{array}{cccc}
   \Delta^{(e_{\text{a}}0)}& 0&\frac{\Omega_{e_{\text{a}}0}}{2} &  0\\
0&   \Delta^{(e_{\text{a}}+)}& \frac{\Omega_{e_{\text{a}}+}}{2}& 0 \\
 \frac{\Omega_{e_{\text{a}}0}^\ast}{2} & \frac{\Omega_{e_{\text{a}}+}^\ast}{2}
 &0 &  \iota_- \frac{\Omega_\mu^\ast}{2} \\
 0& 0 & \iota_-\frac{\Omega_\mu}{2} & 0
\end{array}
\right)\label{adia-03-app}
\end{eqnarray}
which is in the basis of $\{ \lvert \text{leak}, e_{\text{a}}0\rangle  , \lvert \text{leak}, e_{\text{a}}+\rangle,  \lvert e_{\text{a}}\rangle,\lvert \downarrow_{\text{a}}\rangle  \}$. In the equation above, we have added a factor $\iota_-$ in the Rabi frequency because $\iota_-=1$ denotes perfect polarization in the field, while any nonzero $\iota_-$ is related with impure polarization in the field; this is used in the other Hamiltonians in the simulation.

\section{Expansion of DDI with small position fluctuation of molecules}\label{app-expansion}
In this appendix, we study the sensitivity of the DDI with respect to small deviation of the positions of the molecules. Because the interaction is pairwise and the traps in Fig.~\ref{figure-1d} are in the $yz$ plane, we consider two molecules lying in the $yz$ plane as shown in Fig.~\ref{figure-1d}, and the quantization axis is along the $z$-axis. For brevity, we define $\hat{z}\equiv(0, 0, 1) $. The two molecules lie along a vector $\vec{r}$ in the $yz$ plane at an angle $ \mathbf{\theta_{\mathbf{p}}}$ relative to the $z$-axis. If the two molecules are from the three molecules at $O,D_{1,2}$ of Fig.~\ref{figure-1d}, then $ \mathbf{\theta_{\mathbf{p}}}=\vartheta\equiv$ arccos $(1/\sqrt{3})$. By definition, we have $        \vec{r} =  r\mathbf{(0, \, \sin\theta_{\mathbf{p}}, \, \cos\theta_{\mathbf{p}})}$, where we use $\vec{\cdot}$ for a vector so as to avoid confusion in this Appendix. This means that $\hat{r} =  \mathbf{(0, \, \sin\theta_{\mathbf{p}}, \, \cos\theta_{\mathbf{p}})} \equiv (0, s, c)$ where we have defined $c \equiv \cos \mathbf{\theta_{\mathbf{p}}}$ and $s \equiv  \mathbf{\sin\theta_{\mathbf{p}}}$ for brevity. In this case, we can see that $ \hat{z}\cdot\hat{r} = \mathbf{ \cos\theta_{\mathbf{p}}} = c$.

When the two molecules are trapped in the traps, there can be some deviation of their positions with respect to the ideal ones. To characterize this, we define a small vector $\vec{u} = (x, y, z)$ so that the actual two-molecule separation vector is given by
    \begin{equation}
        \vec{R} = \vec{r} + \vec{u} = (x, \, rs + y, \, rc + z).
    \end{equation}

We first consider the interaction between the ancilla and one data molecule, where the interaction is proportional to $  \sin^2\mathbf{\theta_{\mathbf{p}}} e^{-2i\phi}/R^3$ as shown in Eq.~(\ref{ancillaDDI}), where $\phi$ is the azimuthal angle. For the molecules in Fig.~\ref{figure-1d}, we have $\phi=\pm \pi/2$ if $\vec{u}=0$. By expanding to the second order with respect to the small deviation $\vec{u}/r$, we have
\begin{align}
e^{-2i\phi} &\approx -\left[ 1 + \frac{2i}{rs} x \left(1 - \frac{  y}{rs}\right) - \frac{2}{(rs)^2}x ^2 \right] \nonumber \\
&= -1 - \frac{2i}{rs} x + \frac{2}{(rs)^2}x^2 + \frac{2i}{(rs)^2} x y.\label{phi-expansion}
\end{align}

Beside of the azimuthal angle, the interaction can be written as $C\sin^2 \mathbf{\theta_{\mathbf{p}}}/R^3$, i.e.,
\begin{equation}
    J(\vec{R}) = \frac{C}{R^3}\left[1 - (\hat{z}\cdot\hat{R})^2\right] = C \frac{R^2 - (\hat{z}\cdot\vec{R})^2}{R^5} \equiv C \frac{\Xi}{R^5},
\end{equation}
where $C$ is $\frac{\text{\textdong}^2}{8\pi\epsilon_0R^3}$ for the DDI between the ancilla and one data molecule, shown in Eq.~(\ref{ancillaDDI}). We can expand $J(\vec{R})$ up to second order in $\vec{u}/r$, namely,
\begin{eqnarray}
J(\vec{R}) &\approx &\frac{C}{r^3} \left[ s^2 + A_{x,1}\frac{x}{r} + A_{y,1}\frac{y}{r} + A_{z,1}\frac{z}{r} + A_{xx,2}\frac{x^2}{r^2} + A_{yy,2}\frac{y^2}{r^2} + A_{zz,2}\frac{z^2}{r^2} + A_{yz,2}\frac{yz}{r^2} \right]\nonumber\\
    &=&\frac{C}{r^3}  s^2+\mathscr{J}^{(1)}+\mathscr{J}^{(2)},\label{expansion0}
\end{eqnarray}
where $\mathscr{J}^{(1)}$ and $\mathscr{J}^{(2)}$ are the first and second-order terms, respectively. In order to derive the above equation, we first cope with the numerator $\Xi = R^2 - (\hat{z}\cdot\vec{R})^2$. Using $R^2 = r^2 + 2\vec{r}\cdot\vec{u} + u^2$ and $\hat{z}\cdot\vec{R} = rc + z$, the zeroth, first, and second-order terms are
\begin{align}
    \Xi_0 &= r^2(1 - c^2) = r^2 s^2 ,\nonumber\\
    \Xi_1 &= 2r(sy + cz) - 2rc z = 2rs y, \nonumber\\
    \Xi_2 &= u^2 - z^2 = x^2 + y^2. \label{numerator}
\end{align}
Then, we cope with $R^{-5}$. With $\epsilon = \frac{2\hat{r}\cdot\vec{u}}{r} + \frac{u^2}{r^2} = \frac{2(sy + cz)}{r} + \frac{x^2 + y^2 + z^2}{r^2}$, we can similarly have
\begin{equation}
    \frac{1}{R^5} = \frac{1}{r^5}\left[ 1 - 5\frac{\hat{r}\cdot\vec{u}}{r} + \frac{1}{r^2}\left(\frac{35}{2}(\hat{r}\cdot\vec{u})^2 - \frac{5}{2}u^2\right) + \mathcal{O}\left(\frac{u^3}{r^3}\right) \right].\label{denomi}
\end{equation}

By properly combining terms from Eq.~(\ref{numerator}) and Eq.~(\ref{denomi}), the first-order term is
\begin{equation}
    \mathscr{J}^{(1)} = \frac{C}{r^5} \left[  \Xi_1 - 5  \Xi_0 \frac{\hat{r}\cdot\vec{u}}{r} \right] = \frac{C}{r^4} \left[ 2sy - 5s^2(sy + cz) \right] = \frac{C}{r^4} \left[ s(2 - 5s^2)y - 5s^2 cz \right],
\end{equation}
from which one can see that the linear coefficients are given by
\begin{align}
    A_{x,1} &= \mathbf{0} ,\nonumber \\
    A_{y,1} &= s(2 - 5s^2) = \mathbf{\sin\theta_{\mathbf{p}} (2 - 5\sin^2\theta_{\mathbf{p}})} ,\nonumber \\
    A_{z,1} &= -5s^2 c = \mathbf{-5\sin^2\theta_{\mathbf{p}} \cos\theta_{\mathbf{p}}}.
\end{align}

For the second-order terms, we can combine $ \Xi_2$, $-5 \Xi_1\frac{\hat{r}\cdot\vec{u}}{r}$, and $ \Xi_0 \frac{1}{r^2}\left(\frac{35}{2}(\hat{r}\cdot\vec{u})^2 - \frac{5}{2}u^2\right)$ from Eqs.~(\ref{numerator}) and~(\ref{denomi}), so that
\begin{equation}
    \mathscr{J}^{(2)} = \frac{C}{r^5} \left[ (x^2 + y^2) - 10s^2 y^2 - 10sc yz + s^2 \left( \frac{35}{2}(sy + cz)^2 - \frac{5}{2}(x^2 + y^2 + z^2) \right) \right],
\end{equation}
from which we can collect by monomial components. Then, the out-of-plane transverse coefficient $A_{xx,2}$ is
\begin{equation}
    A_{xx,2} = 1 - \frac{5}{2}s^2 =  -\frac{1}{4}(1 - 5\cos 2\mathbf{\theta_{\mathbf{p}})},
\end{equation}
the in-plane radial coefficient $A_{yy,2}$ is
\begin{equation}
    A_{yy,2} = 1 - \frac{25}{2}s^2 + \frac{35}{2}s^4 =  {\frac{1}{16}(21 - 40\cos 2\theta_{\mathbf{p}} + 35\cos 4\theta_{\mathbf{p}})},
\end{equation}
and the in-plane axial coefficient $A_{zz,2}$ is
\begin{equation}
    A_{zz,2} = \frac{5}{2}s^2(7c^2 - 1) = {\frac{5}{16}(3 + 4\cos 2\theta_{\mathbf{p}} - 7\cos 4\theta_{\mathbf{p}})},
\end{equation}
where ``radial'' and ``axial'' do not mean that the direction is along the radial and axial directions of the tweezer, but only refer to the axis of the two-molecule separation axis. Beside the quadratic terms, there is also a cross-term coefficient $A_{yz,2}$, given by
\begin{equation}
    A_{yz,2} = -10sc + 35s^3 c = {5\sin\theta_{\mathbf{p}} \cos\theta_{\mathbf{p}} (7\sin^2\theta_{\mathbf{p}} - 2)}.
\end{equation}
To summarize, the sensitivity coefficients are given in Table~\ref{sensi-table}.

\begin{table}[h!]
\centering
\caption{Sensitivity coefficients for $J(\vec{R}) = \frac{C}{R^3}(1 - (\hat{z}\cdot\hat{R})^2)$ in the $yz$ Plane.}\label{sensi-table}
\begin{tabular}{cccc}
\toprule\hline\hline
\textbf{Term} & \textbf{Coefficient} & \textbf{Trigonometric Form}  & \textbf{Value at} $\theta_{\mathbf{p}}=$arccos$\frac{1}{\sqrt{3}}$ \\ \midrule\hline
$x$ & $A_{x,1}$ & $0$ &0  \\ \addlinespace
$y$ & $A_{y,1}$ & $\sin\theta_{\mathbf{p}} (2 - 5\sin^2\theta_{\mathbf{p}})$ &\(-\dfrac{4\sqrt6}{9}\)  \\ \addlinespace
$z$ & $A_{z,1}$ & $-5\sin^2\theta_{\mathbf{p}} \cos\theta_{\mathbf{p}}$ &  \(-\dfrac{10\sqrt3}{9}\) \\ \midrule\hline
$yz$ & $A_{yz,2}$ & $5\sin\theta_{\mathbf{p}} \cos\theta_{\mathbf{p}} (7\sin^2\theta_{\mathbf{p}} - 2)$ &  \(\dfrac{40\sqrt2}{9}\)\\ \addlinespace
$x^2$ & $A_{xx,2}$ & $-\frac{1}{4}(1 - 5\cos 2\theta_{\mathbf{p}})$  &\(-\dfrac{2}{3}\)\\ \addlinespace
$y^2$ & $A_{yy,2}$ & $\frac{1}{16}(21 - 40\cos 2\theta_{\mathbf{p}} + 35\cos 4\theta_{\mathbf{p}})$  &\(\dfrac{4}{9}\) \\ \addlinespace
$z^2$ & $A_{zz,2}$ & $\frac{5}{16}(3 + 4\cos 2\theta_{\mathbf{p}} - 7\cos 4\theta_{\mathbf{p}})$ & \(\dfrac{20}{9}\)  \\ \bottomrule\hline\hline
\end{tabular}
\end{table}

\begin{table}[h!]
\centering
\caption{Sensitivity coefficients for $J(\vec{R}) = \frac{C}{R^3}(1 -3 (\hat{z}\cdot\hat{R})^2)$ in the $yz$ Plane.}\label{sensi-table2}
\begin{tabular}{ccc}
\toprule\hline\hline
\textbf{Term} & \textbf{Coefficient} & \textbf{Value at} $\theta_{\mathbf{p}}=$arccos$\frac{1}{\sqrt{3}}$ \\ \midrule\hline
$x$ & $A_{x,1} = 0$ &0 \\ \addlinespace
$y$ & $A_{y,1} = 3\sin\theta_{\mathbf{p}}(5\cos^2\theta_{\mathbf{p}} - 1)$ &\(\dfrac{2\sqrt{6}}{3}\)  \\ \addlinespace
$z$ & $A_{z,1} = 3\cos\theta_{\mathbf{p}}(5\cos^2\theta_{\mathbf{p}} - 3)$ &\(-\dfrac{4\sqrt{3}}{3}\)  \\ \midrule\hline
$x^2$ & $A_{xx,2} =  \frac{3}{4}(3 + 5\cos 2\theta_{\mathbf{p}})$ &1  \\ \addlinespace
$y^2$ & $A_{yy,2} = \frac{3}{16}(-3 + 35\cos 4\theta_{\mathbf{p}})$ & \(-\dfrac{17}{3}\)  \\ \addlinespace
$z^2$ & $A_{zz,2} = -\frac{3}{16}(9 + 20\cos 2\theta_{\mathbf{p}} + 35\cos 4\theta_{\mathbf{p}})$ & \(\dfrac{14}{3}\) \\
$yz$ & $A_{yz,2} = 15\cos\theta_{\mathbf{p}} \sin\theta_{\mathbf{p}} (3 - 7\cos^2\theta_{\mathbf{p}})$ &  \(\dfrac{10\sqrt{2}}{3}\)\\ \addlinespace\bottomrule\hline\hline
\end{tabular}
\end{table}

The interaction between two data molecules in this article can be written as
\begin{equation}
    J(\vec{R}) = \frac{C}{R^3}\left(1 - 3(\hat{d}\cdot\hat{R})^2\right) = C \frac{R^2 - 3(\hat{d}\cdot\vec{R})^2}{R^5},
\end{equation}where $C = -\frac{1}{6}\frac{\text{\textdong}^2}{4\pi\epsilon_0R^3}$. One can use the same method as shown above to study the interaction between two data molecules. Because the method is exactly the same, we omit the details, and show the result in Table~\ref{sensi-table2}. In order to compare the results with those in Ref.~\cite{You_2025}, the quadratic second-order coefficients are written as functions of $\cos 2\theta_{\mathbf{p}}$ and $\cos 4\theta_{\mathbf{p}}$, where one should pay attention to the relation between the convention of this article and the one used in Ref.~\cite{You_2025},
\begin{eqnarray}
\text{Ref.~\cite{You_2025}'s }x\ (\text{along }B)
\;&&\longleftrightarrow\;
\text{present }z\ (\text{along }B),\nonumber\\
\text{Ref.~\cite{You_2025}'s }y\ (\text{in-plane transverse})
\;&&\longleftrightarrow\;
\text{present }y\ (\text{in-plane transverse}),\nonumber\\
\text{Ref.~\cite{You_2025}'s }z\ (\text{out-of-plane})
\;&&\longleftrightarrow\;
\text{present }x\ (\text{out-of-plane}).
\end{eqnarray}
Of course, in Ref.~\cite{You_2025} the linear terms are assumed to be irrelevant, which coincides with the conclusion drawn in Sec.~\ref{sec05B1}.

\section{QM-DDI coupling in three molecules of the $\text{C}_2(Z)$ gate}\label{sec-mode-separate}
The QM-DDI coupling between two polar molecules was studied in Refs.~\cite{Lu2026gate,Lu2026gate2}. Here, we consider the case of the ancilla-assisted $\text{C}_2(Z)$ gate, which involves three molecules, one ancilla ``a'' and two data molecules, labeled ``c'' and ``t''. For the setup in Fig.~\ref{figure-1d}, the trap center for the ancilla is at $O(0,0,0)$, while those for trapping the data qubits are at $D_1(0, -L_1\sin\vartheta_1, -L_1\cos\vartheta_1)$ and $D_2(0, L_1\sin\vartheta_1, L_1\cos\vartheta_1)$. The quantization axis is along $\mathbf{z}$.

To be general, we consider that the data molecules are in traps centered at $(0,y_{c0(t0)},z_{c0(t0)})$, covering all the sites $D_j$ with $j\in\{1,\cdots,6\}$ in Fig.~\ref{figure-1d}. Then the Hamiltonian for the three trapped molecules of the gate in Sec.~\ref{sec04} is
\begin{eqnarray}
 \hat{H}_{\text{trap}} &=& \sum_{\alpha=\text{c,t,a}} \sum_{\xi =\text{x,y,z}}\left\{ -\frac{\hbar^2}{2m} \frac{\partial^2}{\partial \xi_\alpha^2 }  +\frac{m\omega_\xi^2\xi_\alpha^2}{2}\right\} + \frac{m\omega_y^2 }{2}\left[(y_{\text{c}}-y_{c0})^2- y_{\text{c}}^2+(y_{\text{t}}-y_{t0})^2- y_{\text{t}}^2 \right] \nonumber\\  && + \frac{m\omega_z^2 }{2}\left[(z_{\text{c}}-z_{c0})^2- z_{\text{c}}^2 + (z_{\text{t}}-z_{t0})^2- z_{\text{t}}^2 \right],\label{H-0}
\end{eqnarray}
where $\hbar$ is the reduced Planck constant, $m$ is the mass of the molecule, $\omega_\xi$ is the angular trap frequency along $\xi\in\{x,y,z\}$, and $\xi_{\alpha}$ is the coordinate of the molecular position along $\xi\in\{x,y,z\}$ for molecule $\alpha\in\{a,c,t\}$. Here, we assume that the traps have the same trap frequencies among the three molecules.

To capture the coupling between the motion and the internal states of the trapped molecules, we define
\begin{eqnarray}
Y_{c(t)}&=& y_{c(t)}-y_{c0(t0)},~~
Z_{c(t)} =  z_{c(t)}-z_{c0(t0)},
\end{eqnarray}
which is small compared to the molecule-molecule separation, and
\begin{eqnarray}
\mathscr{X}_0&=&\frac{1}{\sqrt{3}}(x_{\text{a}} + x_{\text{c}} +x_{\text{t}} )~~
\mathscr{X}_1=\frac{1}{\sqrt{6}}(2x_{\text{a}} - x_{\text{c}} -x_{\text{t}} ),~~
\mathscr{X}_2=\frac{1}{\sqrt{2}}( x_{\text{c}} -x_{\text{t}} ),\nonumber\\
\mathscr{Y}_0&=&\frac{1}{\sqrt{3}}(y_{\text{a}} + Y_{\text{c}} +Y_{\text{t}} )~~
\mathscr{Y}_1=\frac{1}{\sqrt{6}}(2y_{\text{a}} - Y_{\text{c}} -Y_{\text{t}} ),~~
\mathscr{Y}_2=\frac{1}{\sqrt{2}}( Y_{\text{c}} -Y_{\text{t}} ),\nonumber\\
\mathscr{Z}_0&=&\frac{1}{\sqrt{3}}(z_{\text{a}} + Z_{\text{c}} +Z_{\text{t}} )~~
\mathscr{Z}_1=\frac{1}{\sqrt{6}}(2z_{\text{a}} - Z_{\text{c}} -Z_{\text{t}} ),~~
\mathscr{Z}_2=\frac{1}{\sqrt{2}}( Z_{\text{c}} -Z_{\text{t}} ),
\label{transform}
\end{eqnarray}
by which Eq.~(\ref{H-0}) becomes,
\begin{eqnarray}
 \hat{H}_{\text{trap}} &=&  \sum_{\alpha\in\text{0,1,2}} \sum_{\xi \in\{\mathscr{X,Y,Z}\}}\left\{ -\frac{\hbar^2}{2m} \frac{\partial^2}{\partial \xi_\alpha^2 }  +\frac{m\omega_\xi^2\xi_\alpha^2}{2}\right\}, \label{H-3}
\end{eqnarray}
where $\omega_{\xi \in\{\mathscr{X,Y,Z}\}}=\omega_{\xi \in\{{x,y,z}\} }$. Along each direction $\xi\in\{x,y,z\}$, we define a harmonic oscillator length $\ell_\xi =  \sqrt{ \hbar/(m\omega_\xi)}$, and
\begin{eqnarray}
\hat{a}_{\xi\alpha} &=&  \frac{1}{\sqrt{2 } \ell_\xi}\left(\xi_{\alpha}+ \ell_\xi^2 \frac{\partial}{\partial \xi_{\alpha}}\right),
\hat{a}_{\xi\alpha}^\dag=  \frac{1}{\sqrt{2 } \ell_\xi} \left(\xi_{\alpha}- \ell_\xi^2 \frac{\partial}{\partial \xi_{\alpha}}\right), \label{a-b-mode}
\end{eqnarray}
so that Eq.~(\ref{H-0}) can be further written as
\begin{eqnarray}
 \hat{H}_{\text{trap}} &=&  \hbar \sum_{\alpha=0}^2\sum_{\xi\in\{x,y,z\}}\omega_\xi \left(\hat{a}_{\xi\alpha }^\dag \hat{a}_{\xi\alpha}   +\frac{1}{2}\right). \label{H-trap01}
\end{eqnarray}
Similar to Refs.~\cite{Lu2026gate,Lu2026gate2}, $\{\hat{a}_{\xi\alpha }^\dag ,\hat{a}_{\xi\alpha}\}$ are the bosonic creation and annihilation operators along $\xi\in\{x,y,z\}$ for the motional states of the three-molecule motional mode $\alpha\in\{0,1,2\}$, respectively. They are related to $\mathscr{X}_\alpha,~\mathscr{Y}_\alpha,~\mathscr{Z}_\alpha$ via
\begin{eqnarray}
\mathscr{X}_\alpha &=&  \frac{\ell_x}{\sqrt{2}}(\hat{a}_{x\alpha}^\dag + \hat{a}_{x\alpha}),~\mathscr{Y}_\alpha =  \frac{\ell_y}{\sqrt{2}}(\hat{a}_{y\alpha}^\dag + \hat{a}_{y\alpha}),~\mathscr{Z}_\alpha =  \frac{\ell_z}{\sqrt{2}}(\hat{a}_{z\alpha}^\dag + \hat{a}_{z\alpha}).\label{Xtoaadag}
\end{eqnarray}

The DDI between the three molecules couples QM with the internal states, and each interaction arises between two molecules. So, we perform perturbation calculation for the QM-DDI coupling following Refs.~\cite{Lu2026gate,Lu2026gate2}.
\subsection{DDI between the molecules c and t}
The DDI between c and t is given by $-\frac{1}{4\pi\epsilon_0R^3}\frac{\text{\textdong}^2}{6}(1-3\cos^2\theta)$ as shown in Appendix~\ref{app-ddi}. To put it into the form of Table~\ref{sensi-table} and Table~\ref{sensi-table2}, we can write it as
\begin{eqnarray}
\mathscr{J}_{\text{ct}} &=&\frac{C_{\text{ct}}}{R^3} (1-3\cos^2\theta) ,\label{J-fluctuation0-0}
\end{eqnarray}
where $R=2L_1$ if no fluctuation occurs, and
\begin{eqnarray}
C_{\text{ct}} =-\frac{1}{4\pi\epsilon_0}\frac{\text{\textdong}^2}{6} .
\end{eqnarray}
According to Eq.~(\ref{expansion0}) and Table~\ref{sensi-table2}, Eq.~(\ref{J-fluctuation0-0}) can be written as
\begin{eqnarray}
\mathscr{J}_{\text{ct}} &\approx & \frac{1-3\cos^2\vartheta }{(2L_1)^3} C_{\text{ct}}+\frac{C_{\text{ct}}}{(2L_1)^3} \left[  \frac{ A_{y,1} y + A_{z,1} z}{2L_1} + \frac{A_{xx,2} x^2  + A_{yy,2}y^2+ A_{zz,2} z^2 + A_{yz,2} yz}{(2L_1)^2} \right],\nonumber\\\label{app-C01}
\end{eqnarray}
where the first term in the bracket is the first-order term, and the second term is the second-order term. Note that for the geometry in Fig.~\ref{figure-1d}, $\vartheta=$arccos$(1/\sqrt{3})$ and the term $\frac{1-3\cos^2\vartheta }{(2L_1)^3} C_{\text{ct}}$ is zero. According to Eq.~(\ref{transform}), we have
\begin{eqnarray}
x&=& \sqrt{2}\mathscr{X}_2,
\end{eqnarray}
and by Eq.~(\ref{Xtoaadag}), the above equation can be written as
\begin{eqnarray}
x&=& \sqrt{2}\mathscr{X}_2 =   \ell_x (\hat{a}_{x2}^\dag + \hat{a}_{x2}).
\end{eqnarray}
For the other two components, we have
\begin{eqnarray}
y&=& y_{c}-y_{t}-(y_{c0}-y_{t0})=Y_c-Y_t=\sqrt{2}\mathscr{Y}_2=   \ell_y (\hat{a}_{y2}^\dag + \hat{a}_{y2}). \nonumber\\
z&=& z_{c}-z_{t}-(z_{c0}-z_{t0})=Z_c-Z_t=\sqrt{2}\mathscr{Z}_2=   \ell_z (\hat{a}_{z2}^\dag + \hat{a}_{z2}).\label{yz-detail}
\end{eqnarray}
where we should bare in mind that $x,y,$ and $z$ are small compared to the molecule-molecule separation $2L_1$. Insertion of Eq.~(\ref{yz-detail}) into Eq.~(\ref{app-C01}) gives the expansion of $\mathscr{J}_{ct}$ with the small parameters $\ell_{\alpha}/(2L_1)$.

\subsection{DDI between molecules a and c}
QM-DDI coupling between the a and c(t) molecules involves the interaction that changes the total projection of spin. Take the interaction between molecules a and c as an example,
\begin{eqnarray}
\mathscr{J}_{\text{ac}} &=&\frac{C_{\text{ac}}}{R^3} \sin^2\theta e^{-2i\phi} ,\label{J-fluctuation0-1}
\end{eqnarray}
where $R=L_1$ if no fluctuation occurs, and
\begin{eqnarray}
C_{\text{ac}} =\frac{1}{4\pi\epsilon_0}\frac{\text{\textdong}^2}{2} .
\end{eqnarray}
As shown in Eq.~(\ref{phi-expansion}), we have
\begin{align}
e^{-2i\phi} \approx  -1 - \frac{2i}{L_1\sin\vartheta} x + \frac{2}{(L_1\sin\vartheta)^2}x^2 + \frac{2i}{(L_1\sin\vartheta)^2} x y.\label{phi-expansion2}
\end{align}
And with Eq.~(\ref{expansion0}) and Table~\ref{sensi-table}, Eq.~(\ref{J-fluctuation0-1}) can be written as
\begin{eqnarray}
\mathscr{J}_{\text{ac}} &\approx &\left\{ \frac{\sin^2\vartheta }{ L_1^3} C_{\text{ac}}+\frac{C_{\text{ac}}}{L_1^3} \left[  \frac{ A_{y,1} y + A_{z,1} z}{L_1} + \frac{A_{xx,2} x^2  + A_{yy,2}y^2+ A_{zz,2} z^2 + A_{yz,2} yz}{L_1^2} \right]\right\}e^{-2i\phi},\nonumber\\\label{app-C03}
\end{eqnarray}
where for the molecules a and c, we have
\begin{eqnarray}
 x&=& \frac{\sqrt{6}\mathscr{X}_1-\sqrt{2}\mathscr{X}_2 }{2}=\frac{\ell_x}{2}\left[ \sqrt{3}  (\hat{a}_{x1}^\dag + \hat{a}_{x1}) -    (\hat{a}_{x2}^\dag + \hat{a}_{x2}) \right].\label{app-ac-01}
\end{eqnarray}
Similarly, we have
\begin{eqnarray}
y&=&   \frac{\ell_y}{2}\left[ \sqrt{3}  (\hat{a}_{y1}^\dag + \hat{a}_{y1}) -    (\hat{a}_{y2}^\dag + \hat{a}_{y2}) \right],\nonumber\\
z&=&   \frac{\ell_z}{2}\left[ \sqrt{3}  (\hat{a}_{z1}^\dag + \hat{a}_{z1}) -    (\hat{a}_{z2}^\dag + \hat{a}_{z2}) \right].\label{app-ac-02}
\end{eqnarray}
Using Eqs.~(\ref{phi-expansion2}),~(\ref{app-ac-01}), and~(\ref{app-ac-02}), Eq.~(\ref{app-C03}) can be cast into a form in terms of the creation and annihilation operators.

\subsection{DDI between molecules a and t}
Because of the definition in Eq.~(\ref{transform}), the interaction between the ancilla and the molecule t is different from that between the ancilla and the molecule c. Equations.~(\ref{app-ac-01}) and~(\ref{app-ac-02}) should be
\begin{eqnarray}
 x&=&  \frac{\ell_x}{2}\left[ \sqrt{3}  (\hat{a}_{x1}^\dag + \hat{a}_{x1}) +    (\hat{a}_{x2}^\dag + \hat{a}_{x2}) \right],\nonumber\\
y&=&   \frac{\ell_y}{2}\left[ \sqrt{3}  (\hat{a}_{y1}^\dag + \hat{a}_{y1}) +    (\hat{a}_{y2}^\dag + \hat{a}_{y2}) \right],\nonumber\\
z&=&   \frac{\ell_z}{2}\left[ \sqrt{3}  (\hat{a}_{z1}^\dag + \hat{a}_{z1}) +    (\hat{a}_{z2}^\dag + \hat{a}_{z2}) \right].\label{app-ac-03}
\end{eqnarray}
for the case of the interaction between molecules a and t. In this case, Eqs.~(\ref{phi-expansion2}) and (\ref{app-ac-03}) can be inserted into Eq.~(\ref{app-C03}), so that we will have the DDI in terms of the creation and annihilation operators.

\subsection{Gate fidelity with motional state}\label{app-fidelity}
The gate with the motional state has a fidelity defined in a way that should take into account the change of motional state, i.e., should include error from the entanglement between the internal states and the motional states.

Note that in Eq.~(\ref{adia-5}), the gate map is written only for the data molecules, but in practice the input is with the ancilla. If not including the motional state, the gate map in the basis $\{\lvert\downarrow_{\text{a}},\uparrow,\uparrow\rangle, \lvert\downarrow_{\text{a}},\uparrow,\downarrow\rangle, \lvert\downarrow_{\text{a}},\downarrow,\uparrow\rangle, \lvert\downarrow_{\text{a}},\downarrow,\downarrow\rangle\}$ should be
\begin{eqnarray}
U &=& \left(
  \begin{array}{cccc}
    1& 0 & 0&0\\
    0 & 1 &0&0\\
    0 &0 & 1&0\\
    0& 0 & 0&e^{i\varphi}\\
    \end{array}
  \right) ,\label{cnot}
  \end{eqnarray}
where $\varphi$ is from the phase of the second adiabatic microwave pulse. For the case of the $C_{2}(Z)$ gate, one $2\pi$ pulse is sufficient. During the pulse, the input states $\{\lvert\downarrow_{\text{a}},\uparrow,\uparrow\rangle, \lvert\downarrow_{\text{a}},\uparrow,\downarrow\rangle, \lvert\downarrow_{\text{a}},\downarrow,\uparrow\rangle\}$ will evolve according to the corresponding Hamiltonian. Because of the diabatic error, the gate map becomes,
\begin{eqnarray}
 \mathscr{U} &=& \left(
  \begin{array}{cccc}
    a& 0 & 0&0\\
    0 & b &0&0\\
    0 &0 & c&0\\
    0& 0 & 0&-1\\
    \end{array}
  \right) ,\label{realgate01}
  \end{eqnarray}
where $a,b,c$ are the inner products between $\{\lvert\downarrow_{\text{a}},\uparrow,\uparrow\rangle, \lvert\downarrow_{\text{a}},\uparrow,\downarrow\rangle, \lvert\downarrow_{\text{a}},\downarrow,\uparrow\rangle\}$ and the wavefunctions calculated by the corresponding Hamiltonian composed of the microwave driving and DDI. The input state $\lvert\downarrow_{\text{a}},\downarrow,\downarrow\rangle$ does not involve DDI, and can attain perfect Rabi oscillation in principle. Then, the fidelity is given by~\cite{Pedersen2007}
\begin{eqnarray}
 \mathcal{F}_{\text{2}}&=&  \frac{1}{20}\left[  |\text{Tr}(U^\dag \mathscr{U})|^2 + \text{Tr}(U^\dag \mathscr{U}\mathscr{U}^\dag U ) \right], \label{fidelityError01}
\end{eqnarray}
and the infidelity is given by $\mathcal{E}=1- \mathcal{F}_{\text{2}}  .$

If motional state is involved, we can consider an input state of a general superposition
\begin{eqnarray}
\lvert \psi\rangle= (
\alpha_1\lvert\downarrow_{\text{a}},\uparrow,\uparrow\rangle+\alpha_2 \lvert\downarrow_{\text{a}},\uparrow,\downarrow\rangle+\alpha_3  \lvert\downarrow_{\text{a}},\downarrow,\uparrow\rangle+\alpha_4 \lvert\downarrow_{\text{a}},\downarrow,\downarrow\rangle)\otimes \lvert \mathcal{M}\rangle,
\label{initialM}
  \end{eqnarray}
where $\lvert \mathcal{M}\rangle$ is the initial motional state. The desired final state should be
\begin{eqnarray}
\lvert \psi(T)\rangle_{\text{ideal}}=   (
\alpha_1\lvert\downarrow_{\text{a}},\uparrow,\uparrow\rangle+\alpha_2 \lvert\downarrow_{\text{a}},\uparrow,\downarrow\rangle+\alpha_3  \lvert\downarrow_{\text{a}},\downarrow,\uparrow\rangle+e^{i\varphi}\alpha_4 \lvert\downarrow_{\text{a}},\downarrow,\downarrow\rangle)   \otimes \lvert \mathcal{M}'\rangle,
\label{initialM2}
  \end{eqnarray}
  according to Eq.~(\ref{cnot}).

  In Eq.~(\ref{initialM2}), the final motional state $\lvert \mathcal{M}'\rangle$ is the state of the in-trap motion. The gate only concerns the internal state so that $\lvert \mathcal{M}'\rangle$ should not necessarily be equal to the initial one as long as there is no entanglement between the internal and the motional states. Note that this is very similar to that an overall phase to a quantum state is trivial.

With the understanding of the essence of the gate at hand, the gate fidelity can be studied. Due to the diabatic leakage, the gate actually results in a final state
\begin{eqnarray}
\lvert \psi(T)\rangle_{\text{real}}&=&
\alpha_1\lvert\downarrow_{\text{a}},\uparrow,\uparrow\rangle\otimes \lvert \mathcal{M}_1\rangle+ \alpha_2 \lvert\downarrow_{\text{a}},\uparrow,\downarrow\rangle\otimes \lvert \mathcal{M}_2\rangle +\alpha_3  \lvert\downarrow_{\text{a}},\downarrow,\uparrow\rangle\otimes \lvert \mathcal{M}_3\rangle\nonumber\\
&& +e^{i\varphi}\alpha_4 \lvert\downarrow_{\text{a}},\downarrow,\downarrow\rangle \otimes \lvert \mathcal{M}\rangle,
\label{initialM3}
  \end{eqnarray}
where $\lvert \mathcal{M}_{1,2,3}\rangle$ can differ from the initial state $\lvert \mathcal{M}\rangle$.
To quantify the quality of the gate, one can see that if we impose $\lvert \mathcal{M}'\rangle=\lvert \mathcal{M}\rangle$, then Eq.~(\ref{realgate01}) should be updated to
\begin{eqnarray}
 \mathscr{U} &=& \left(
  \begin{array}{cccc}
    a\text{Tr}\langle\mathcal{M}| \mathcal{M}_1\rangle& 0 & 0&0\\
    0 &  b\text{Tr}\langle\mathcal{M}| \mathcal{M}_2\rangle &0&0\\
    0 &0 & c\text{Tr}\langle\mathcal{M}| \mathcal{M}_3\rangle&0\\
    0& 0 & 0&-1\\
    \end{array}
  \right) .\label{realgate01-2}
  \end{eqnarray}
Equation~(\ref{realgate01-2}) is a simplified version of the more involved one. This is because one should calculate the inner product in the equation without separating the wavefunction of the internal part from the in-trap motion. In numerics, what we do is to calculate the actual wavefunction of the whole system, where the actual internal wavefunction has several state components, each with their own in-trap motional state.

The procedure above is for the two-qubit gates. The case of three and four-qubit gates can be studied in a similar way. Note that in the three-qubit gate, though there are only four classes of different input states, which are classified according to the DDI, the actual simulation should sample more than one input states within each class as long as the sets of DDI differ between states. Similarly, for the four-qubit gate, the simulation should carefully treat each input state as long as it possesses a unique set of DDI among the molecules.  

\end{appendix}

%


\end{document}